\documentclass[acmsmall]{acmart}

\usepackage{amsmath,amssymb,amsfonts}
\usepackage{algorithmic}
\usepackage{graphicx}
\usepackage{textcomp}
\usepackage{xcolor}
\usepackage{balance}
\usepackage{enumitem}
\usepackage{booktabs}
\usepackage{multirow}
\usepackage{multicol,makecell}
\usepackage{xspace}
\usepackage{graphicx}
\usepackage{xcolor}
\usepackage{fancyhdr}
\usepackage[normalem]{ulem}
\usepackage[bottom]{footmisc}
\usepackage{courier}
\usepackage{dblfloatfix}
\usepackage{longfbox}
\usepackage{caption}
\usepackage{xurl}                  %
\usepackage[]{hyperref}
\usepackage[colorinlistoftodos,prependcaption,textsize=scriptsize]{todonotes} %

\usepackage[sort&compress]{cleveref}
\usepackage{xstring}               %
\usepackage{pifont}   %
\usepackage{xparse}      %
\usepackage{expl3}       %
\usepackage{tikz}
\usepackage{comment}
\usepackage[most]{tcolorbox}
\usepackage{kotex}

\definecolor{pred}{rgb}{0.7843, 0.0039, 0.3137} %
\definecolor{darknavy}{rgb}{0, 0, 0.8}
\definecolor{darkgreen}{rgb}{0, 0.5, 0}
\definecolor{blue(munsell)}{rgb}{0.0, 0.5, 0.69}
\definecolor{darkorange}{rgb}{0.80, 0.44, 0.00}

\newcounter{obscnt}
\renewcommand{\theobscnt}{\arabic{obscnt}}

\crefname{obscnt}{Observation}{Observations}
\Crefname{obscnt}{}{}

\newcommand{\observation}[2]{%
  \par\addvspace{-2pt}%
  \refstepcounter{obscnt}%
  \label{#1}%
  \begin{tcolorbox}[enhanced,
    breakable,                    %
    colback=lightgray!20!white,   %
    boxrule=0pt,                  %
    arc=4pt,                      %
    left=0pt,right=0pt,top=0.1pt,bottom=0pt %
  ]%
    \head{Observation~\theobscnt}\space{#2}
  \end{tcolorbox}%
  \par\addvspace{-2pt}%
  \noindent\ignorespaces
}

\newcounter{take}
\newcommand{\take}[1]{%
  \par\addvspace{-2pt}%
  \refstepcounter{take}%
  \begin{tcolorbox}[enhanced,
    breakable,                    %
    colback=lightgray!35!white,   %
    boxrule=0.5pt,                  %
    arc=4pt,                      %
    left=0pt,right=0pt,top=0.1pt,bottom=0pt %
  ]%
    \head{Takeaway~\thetake}\space{#1}%
  \end{tcolorbox}%
  \par\addvspace{-2pt}%
  \noindent\ignorespaces
}

\ExplSyntaxOn
\cs_new_protected:Npn \__ssd_make:n #1
  {
    \tl_set:Nn \l_tmpa_tl { #1 }
    \tl_remove_all:Nn \l_tmpa_tl { ~ } %
    \tl_replace_all:Nnn \l_tmpa_tl {A-samsung980}   {\allowbreak\mbox{-A}}
    \tl_replace_all:Nnn \l_tmpa_tl {B-wd850}        {\allowbreak\mbox{-B}}
    \tl_replace_all:Nnn \l_tmpa_tl {C-solidigmP41}  {\allowbreak\mbox{-C}}
    \tl_replace_all:Nnn \l_tmpa_tl {D-samsungPM9A1} {\allowbreak\mbox{-D}}
    \tl_replace_all:Nnn \l_tmpa_tl {E-kingstonFR}   {\allowbreak\mbox{-E}}
    \tl_replace_all:Nnn \l_tmpa_tl {F-crucialP3}    {\allowbreak\mbox{-F}}
    \tl_replace_all:Nnn \l_tmpa_tl {G-hynixP41}     {\allowbreak\mbox{-G}}
    \tl_replace_all:Nnn \l_tmpa_tl {H-kioxiaXG8}    {\allowbreak\mbox{-H}}
    \tl_replace_all:Nnn \l_tmpa_tl {I-micron2400}   {\allowbreak\mbox{-I}}
    \tl_replace_all:Nnn \l_tmpa_tl {J-crucialT500}  {\allowbreak\mbox{-J}}
    \tl_replace_all:Nnn \l_tmpa_tl {K-seagateFC530} {\allowbreak\mbox{-K}}
    \tl_replace_all:Nnn \l_tmpa_tl {L-hp900}        {\allowbreak\mbox{-L}}
    \tl_replace_all:Nnn \l_tmpa_tl {M-solidigmP44}  {\allowbreak\mbox{-M}}
    \tl_replace_all:Nnn \l_tmpa_tl {N-hp700}        {\allowbreak\mbox{-N}}
    \tl_replace_all:Nnn \l_tmpa_tl {O-micron3400}   {\allowbreak\mbox{-O}}
    \tl_replace_all:Nnn \l_tmpa_tl {, } {/}
    SSD\l_tmpa_tl                     %
  }
\NewDocumentCommand \ssdlist { m }
  { \__ssd_make:n { #1 } \xspace }
\ExplSyntaxOff

\newcommand{\samsung}{{Mfr.~A}\xspace}
\newcommand{\toshiba}{{Mfr.~B}\xspace}
\newcommand{\solidigm}{{Mfr.~C}\xspace}
\newcommand{\micron}{{Mfr.~D}\xspace}
\newcommand{\hynix}{{Mfr.~E}\xspace}
\newcommand{\kioxia}{{Mfr.~F}\xspace}
\newcommand{\ymtc}{{Mfr.~G}\xspace}

\newcommand{\typeA}{Type\allowbreak\mbox{-1}\xspace}
\newcommand{\typeB}{Type\allowbreak\mbox{-2}\xspace}

\newcommand{\attackA}{CAB\xspace}
\newcommand{\AttackA}{Charge-and-Burst\xspace}

\newcommand{\attackB}{WSF\xspace}
\newcommand{\AttackB}{Weak-Spot Focus\xspace}

\newcommand{\webserver}{\textsf{web server}\xspace}
\newcommand{\videoserver}{\textsf{video server}\xspace}

\newcommand{\RH}{\ensuremath{_{\textsf{RH}}}\xspace}

\newcommand\minorm[1]{{\color{black}{#1}}}

\newcommand{\fig}[1]{{Fig.~#1}\xspace}
\newcommand{\figs}[1]{{Figs.~#1}\xspace}
\newcommand{\sect}[1]{{\S#1}\xspace}
\newcommand{\tbl}[1]{{Table~#1}\xspace}

\newcommand\inum[1]{(\textit{#1})\xspace}
\newcommand{\head}[1]{{\noindent\textbf{#1.}\xspace}}
\newcommand{\vth}{V$_{\text{TH}}$\xspace}
\newcommand{\vref}{V$_{\text{REF}}$\xspace}

\newcommand{\vpass}{V$_{\text{PASS}}$\xspace}

\newcommand{\usec}{\textmu{}s\xspace}

\newcommand{\degreec}[1]{#1$^\circ$C\xspace}
\newcommand{\mytilde}{\textasciitilde}

\DeclareRobustCommand\bcirc[1]{\tikz[baseline=(char.base)]{
           \node[shape=circle,draw,inner sep=0pt,fill=black, text=white] (char) {#1};}}
\DeclareRobustCommand\wcirc[1]{\tikz[baseline=(char.base)]{
           \node[shape=circle,draw,inner sep=0pt,fill=white] (char) {#1};}}

\newcommand{\yellowdiamond}{%
    \tikz[baseline=0.0ex,scale=2]
    \draw[fill=orange, draw=black, line width=0.3pt]
      (0,0) -- (0.15em,0.15em) -- (0,0.3em) -- (-0.15em,0.15em) -- cycle;}

\AtBeginDocument{%
  }

\acmConference[POMACS '26]{the ACM on Measurement and Analysis of Computing Systems}{June 08--12, 2026}{Ann Arbor, MI}
\acmJournal{POMACS}
\acmYear{2026} \acmVolume{10} \acmNumber{1} \acmArticle{10}
\acmMonth{3} \acmPrice{} \acmDOI{10.1145/3788092}

\authorsaddresses{}
\setcopyright{none}
\makeatletter
\renewcommand\footnotetextcopyrightpermission[1]{}
\makeatother
\makeatletter
\AtBeginDocument{%
  \fancypagestyle{standardpagestyle}{%
    \fancyfoot{}%
  }%
}
\makeatother

\begin{document}

\setcounter{section}{0}
\title{Experimental Study on System-Level Performance Impact of Read Disturbance in Modern SSDs}
\newpage
\author{Yonggon Park}
\email{nanimdo@postech.ac.kr}
\orcid{0009-0008-6085-9200}
\affiliation{%
  \institution{POSTECH}
  \country{Republic of Korea}
}

\author{Hyunuk Cho}
\email{gusdnr9779@postech.ac.kr}
\orcid{0000-0002-1554-1884}
\affiliation{%
  \institution{POSTECH}
  \country{Republic of Korea}
}

\author{Onur Mutlu}
\email{omutlu@gmail.com}
\orcid{0000-0002-0075-2312}
\affiliation{%
  \institution{ETH Zurich}
  \country{Switzerland}
}

\author{Sungjin Lee}
\email{sungjin.lee@postech.ac.kr}
\orcid{0000-0002-9753-2286}
\affiliation{%
  \institution{POSTECH}
  \country{Republic of Korea}
}

\author{Jisung Park}
\email{jisung.park@postech.ac.kr}
\orcid{0000-0002-1826-9003}
\affiliation{%
  \institution{POSTECH}
  \country{Republic of Korea}
}

\renewcommand{\shortauthors}{Park et al.}
\setcounter{page}{1}
\pagestyle{standardpagestyle}

\renewcommand{\shortauthors}{Yonggon Park, Hyunuk Cho, Onur Mutlu, Sungjin Lee, and Jisung Park}

\begin{abstract}
This work investigates the system-level performance impact of read disturbance in modern NAND flash-based SSDs, aiming to provide new insights that can help develop better storage architectures and optimize system software.
Continuous improvement in storage density over decades has led NAND flash memory to play a vital role in modern computing systems, but it also comes at a cost of significant reliability degradation.
Among various error sources, read disturbance has gained growing attention as a major reliability concern due to its rapidly increasing impact, which can significantly affect system I/O performance by exacerbating SSD-internal reliability-management overheads.
Although a large body of prior work has focused on device-level characterizations and optimizations, the system-level performance impact of read disturbance still remains largely uninvestigated.
To address this gap, this work conducts a rigorous experimental study using 15 modern NVMe SSDs from 10 major vendors in two ways.
First, we comprehensively analyze the system-level performance impact of read disturbance under diverse workloads and operating conditions.
Second, to highlight the importance of efficient read-disturbance management, we showcase a new possible SSD-performance attack as a case study, demonstrating that an adversary can significantly degrade the I/O performance of other concurrently running processes by exploiting read disturbance alone in commodity SSDs.
Based on our experimental study, we make 16 new observations and 7 takeaway lessons, which lead to 6 key directions for future improvements at the host-system and SSD-architecture levels to better cope with read disturbance.
\end{abstract}


\maketitle
\section{Introduction} \label{sec:intro} \thispagestyle{empty}

Owing to the continuous and significant improvements in storage density over decades, NAND flash memory has become the predominant memory technology for modern storage systems. 
Since 2014, the storage density of NAND flash memory has almost doubled every two years~\cite{choi-isscc-2014, jung-isscc-2024}, which has enabled modern solid-state drives (SSDs) to offer unprecedented single-device capacities (e.g., 128~TB~\cite{samsung128TB}) at a lower cost per bit.
Three key technologies have driven such advancement: \inum{i}~aggressive process scaling, which packs more cells into the same die area, \inum{ii}~multi-level cell (MLC) techniques, which allow each cell to store multiple bits (e.g., triple-level cell (TLC)), and \inum{iii}~3D stacking, which vertically integrates hundreds of wordline (WL) layers in a NAND flash die.

Unfortunately, the continuous storage-density improvements come at a cost of significant reliability degradation in modern NAND flash memory, aggravating the reliability impact of various error sources, such as read disturbance~\cite{ha-tcad-2016, liu-asplos-21, chun-cal-2025, xiong-tos-2018, ren-nvmsa-2023, cai-dsn-2015, zambelli-irps-17, meng-fms-19}, program interference~\cite{park-dac-2016, kim-dac-2017, cai-hpca-2017, cai-iccd-2013}, and retention loss~\cite{mizoguchi-imw-2017, yang-tcad-2025, liu-target-2012,luo-acm-2018}.
NAND flash memory stores data based on a cell's threshold voltage (\vth) level that highly depends on the amount of charge inside the cell.
The unique cell designs and organizations of 3D NAND flash memory render flash cells more susceptible to charge leakage~\cite{mizoguchi-imw-2017, luo-acm-2018} and circuit-level disturbance~\cite{chun-cal-2025, ren-nvmsa-2023, xiong-tos-2018}.
Storing more bits in a smaller cell reduces the margin between different \vth levels, drastically increasing the reliability impact of \vth shifts caused by the error sources~\cite{frickey-irps-2024}.

Among the various error sources, \emph{read disturbance} has gained increasing attention as a major reliability concern in modern NAND flash memory.
To read a page from a target WL, NAND flash memory needs to apply high pass-through voltage \vpass (e.g., 6~V) to \emph{all} non-target WLs in the same block, which unintentionally programs the WLs slightly.
Such disturbance can cause data corruption (i.e., a permanent data loss) when a block experiences a large number of page reads, e.g., 400K~\cite{chun-cal-2025}.
The disturbance effect of a single page read significantly increases in modern NAND flash memory due to two key reasons. 
First, reading a page from a WL requires more sensing operations in advanced MLC techniques, exposing non-target cells to high \vpass for a longer time.
Second, a single page read disturbs more pages, as the number of pages per block increases rapidly.

The increasing reliability impact of read disturbance, in turn, can significantly affect the performance of modern SSDs, exacerbating the overheads of SSD-internal tasks, such as \emph{read retry}~\cite{park-asplos-2021, chun-hpca-2024} and \emph{read reclaim}~\cite{ha-tcad-2016, liu-asplos-21, zhang-tcad-2022}.
When the SSD controller fails to correct all the raw bit errors in a read page with error-correcting codes (ECC), it \emph{retries} reading the page with adjusted read-reference voltage (\vref) levels until the page's raw bit-error rate (RBER) decreases below the ECC's error-correction capability.
Before read retry becomes unable to sufficiently reduce a page's RBER due to excessive read disturbance, the SSD controller needs to perform read reclaim that removes the page's transient errors by rewriting its data to another free page.
Even though both read retry and read reclaim are essential to ensuring the reliability of stored data, their additional read and write operations can significantly affect SSD performance~\cite{park-asplos-2021,ye-asplos-2024,chun-cal-2025}.
To cope with the increasing reliability impact of read disturbance, modern SSDs need to invoke read retry and read reclaim more frequently, aggravating their performance overheads.

\textbf{Our goal} in this work is to provide new insights into the system-level performance impact of read disturbance in modern NAND flash-based SSDs, which has not yet been thoroughly investigated in the literature, but can help develop better storage architectures and optimize system software.
To this end, we conduct a rigorous experimental study using 15 modern NVMe SSDs~\cite{samsung980pro, samsungpm9a1, wdblacksn850pro, seagatefirecuda530, micron3400, kingstonfuryrenegade, hynixp41, crucialp3plus, micron2400, solidigmp41plus, hpfx900pro, solidigmp44pro, kioxiaxg8, crucialt500, hpfx700} from 10 major vendors in two ways.
First, we analyze the system-level performance impact of read disturbance under diverse workloads and operating conditions to \inum{i}~evaluate the effectiveness of existing reliability-management techniques in modern SSDs, \inum{ii}~identify their limitations, and \inum{iii}~explore new optimization opportunities.
Second, as a case study, we showcase a new possible SSD-performance attack that can severely degrade the I/O performance of other applications sharing the same SSD by causing excessive overheads for read-disturbance management, highlighting the importance of efficient read-disturbance management in modern storage systems.

Based on our experimental study, we make \minorm{16} key observations and \minorm{7} takeaway lessons, which lead to 6 key directions for future improvements at the host-system and SSD-architecture levels to better cope with read disturbance.
We highlight four observations that are especially important.
First, read disturbance can significantly degrade SSD performance, e.g., lowering read bandwidth by more than 79.1\% for 32.7 seconds, even under sequential-read workloads that are commonly considered best for SSD performance in prior work~\cite{hu-tc-2013,mao-tcad-2018,ghiasi-asplos-2022}.
Second, severe and consistent performance fluctuations can occur for a long time (e.g.,~several hours) in modern SSDs when a number of blocks trigger read-disturbance management tasks simultaneously.
Third, the performance impact of read disturbance highly depends on SSD-internal management, significantly varying even across SSDs using the same NAND flash chips.
Fourth, an application's I/O performance can significantly degrade due to read disturbance caused by other applications sharing the same SSD.

Our case study experimentally demonstrates a new possible performance attack that can cause significant I/O slowdowns to other concurrent processes by exploiting read disturbance.
The attack issues only sequential reads within its own exclusive logical-block address (LBA) range \emph{without} requiring direct access to the target application's data, any host-privilege escalation, or more CPU cores than the target.
Our evaluation shows that the attack can \inum{i}~burst significant performance fluctuations within a short time window and/or \inum{ii}~sustain long I/O slowdowns, which severely affect both the I/O bandwidth and read tail latency of other applications that share the same SSD.   

This work makes the following key contributions:
\begin{itemize}[leftmargin=*, noitemsep, topsep=0pt]
    \item We present the first rigorous study on the system-level performance impact of read disturbance by testing 15 commodity NVMe SSDs under various operating scenarios, which leads to \minorm{16} new observations that highlight the importance of efficient read-disturbance management.
    \item We propose a new possible SSD-performance attack that can severely degrade a target application's I/O performance by exacerbating read-disturbance management overheads, experimentally demonstrating the attack's feasibility and effectiveness in commodity SSDs.
    \item Based on our new observations, we present six future improvements to better cope with read disturbance that becomes more severe as the storage density of NAND flash memory increases.
\end{itemize}

\section{Background}\label{sec:background}
We provide a brief background on modern SSDs necessary to understand the rest of the paper.
\subsection{NAND Flash-Based SSDs}\label{ssec:bg_nand_flash}

\begin{figure}[b]
     \centering
     \includegraphics[width=0.98\linewidth]{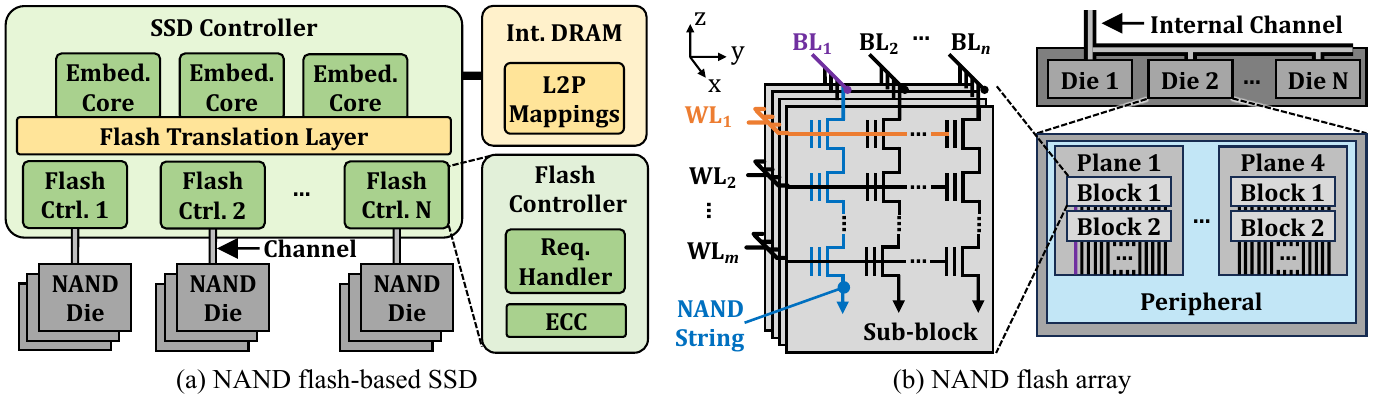}
     \caption{Overview of modern NAND flash-based SSD.}
     \label{fig:nand_ssd}
\end{figure}

\fig{\ref{fig:nand_ssd}(a)} shows an organizational overview of modern NAND flash-based SSDs that commonly consist of three key components: \inum{i}~NAND flash array, \inum{ii}~SSD controller, and \inum{iii}~internal DRAM.

\head{NAND Flash Array}
\fig{\ref{fig:nand_ssd}(b)} illustrates the hierarchical organization of 3D NAND flash memory.
A group of vertically stacked flash cells are serially connected, which is called a \emph{NAND string}.
Each NAND string is connected to a bitline (BL), and NAND strings in the same $y$-$z$ plane constitute a \emph{sub-block}, where a single wordline (WL) connects all flash cells at the same vertical location.
A \emph{block} comprises multiple sub-blocks (e.g., 2 to 8) of which WLs at the same vertical location are connected and thus share the same WL voltage.
A \emph{plane} consists of hundreds of blocks that share all BLs in the plane.
A \emph{die} (or a \emph{chip}) contains several planes (e.g., 4 or 6), and multiple dies can be packaged together while sharing the package's command and data bus, i.e., a \emph{channel}. 

Modern SSDs contain a large number of dies that can concurrently operate (e.g., 32 dies~\cite{kang-isscc-2019}), enabling high internal parallelism.
To enhance random-read performance, manufacturers have recently developed the independent-plane read~\cite{higuchi-isscc-2021} that allows multiple planes in the same die to perform page reads independently of each other.
Despite the high die- and plane-level parallelisms, SSD performance is often bottlenecked by the channel transfer rate (e.g., 1.2--2.4~GB/s~\cite{kang-isscc-2019, kim-isscc-2022, onfi-5-0}) due to the limited number of channels (e.g., 8 or 16~\cite{samsungpm-pm1743-16channels}) that connect the dies to the SSD controller.

\head{SSD Controller}
An SSD controller consists of \inum{i}~multiple embedded cores and \inum{ii}~hardware flash controllers.
The cores run SSD firmware, called \emph{flash translation layer} (FTL), which performs various internal management tasks, such as address translation~\cite{gupta-asplos-2009, jiang-msst-2011, sun-asplos-2023}, garbage collection~\cite{lee-tcad-2013, yang-hotstorage-2019, kang-cm-2017, guo-ipdps-2017, choi-hpdc-2018, shahidi-sc-2016, kang-dac-2018, cui-date-2018, lee-ispass-2011}, reliability management~\cite{cai-hpca-2017, chun-cal-2025, zhang-tcad-2022, park-dac-2016, luo-acm-2018, liu-target-2012}, wear leveling~\cite{dh-tcad-2022, li-msst-2019, murugan-msst-2011}, request scheduling~\cite{tavakkol-isca-2018, jun-nvmsa-2015, jung-hpca-2014, elyasi-asplos-2017, wu-fast-2012}, etc.
A flash controller is in charge of request handling, data randomization, and error-correcting codes (ECC) for the associated NAND flash dies.
Modern SSDs typically employ 8 to 16 \emph{per-channel} flash controllers to prevent ECC from delaying data transfer through the channels~\cite{micheloni-inside-ssd-2018, xue-iccd-2022, tavakkol-fast-2018}.

\head{Internal DRAM}
Modern SSDs employ large low-power DRAM (e.g., several GBs~\cite{samsung-990pro}) for metadata store and data buffer.
Except for some cost-optimized SSDs~\cite{zhang-dramless-2020, micron2400, solidigmp41plus}, it is common to keep all logical-to-physical (L2P) address mappings in internal DRAM to avoid expensive NAND flash accesses for address translation. 
Doing so requires 0.1\% of the SSD capacity to support 4-KiB mapping granularity~\cite{gupta-asplos-2009, zhang-fast-2012}, which consumes most of the internal DRAM capacity.
FTL also needs to maintain various metadata for reliability management, such as the program and erase (P/E)-cycle count, retention age, and read count for each block.
A small fraction of internal DRAM (e.g., tens of MBs~\cite{hwang-atc-2024}) is used as a write buffer to hide the long write latency of NAND flash memory~\cite{kim-fast-2008, lee-islped-2021}.

\subsection{NAND Flash Operations}
\label{ssec:bg_nand_oprs}
Three basic operations enable access to NAND flash memory: \inum{i}~program, \inum{ii}~erase, and \inum{iii}~read.

\head{Program and Erase Operations}
A flash cell stores data using its threshold voltage (\vth) level that can change in a nonvolatile manner depending on the amount of charge in the cell's floating gate or charge trap.
A program operation injects electrons into target cells at a page granularity (e.g., 16 KiB), increasing their \vth levels.
An erase operation ejects electrons from target cells at a block granularity, which decreases their \vth levels.
Injecting/ejecting electrons into/from a large number of cells requires applying a high voltage (e.g., $>$20~V) to the cells for a long time (e.g., 400~\usec and 3.5~ms for program and erase operations, respectively), which physically damages target cells.

\head{Read Operation}
NAND flash memory reads stored data from target cells at a page granularity by identifying whether current flows through the corresponding NAND strings~\cite{micheloni-inside-nand-2010, park-asplos-2021, park-micro-2022}.
When applying read-reference voltage \vref to a target WL, the cells connected to the WL operate as either a resistor (if \vth{}$<$\vref{}) or an open switch (if \vth{}$>$\vref{}).
A NAND flash chip also applies pass voltage \vpass to other WLs in the same block which is high enough (e.g., 6~V) to make all non-target cells operate as a resistor regardless of their \vth levels. 
Doing so ensures that only a target cell's \vth level dictates the corresponding NAND string's conductance.
If the NAND string conducts flow, the sensing logic interprets the cell's data as `\texttt{1}'; otherwise, it reads as `\texttt{0}'.

\subsection{Reliability Issues in Modern SSDs}
\label{ssec:bg_nand_rel}
\begin{figure}[t]
     \centering
     \includegraphics[width=0.83\linewidth]{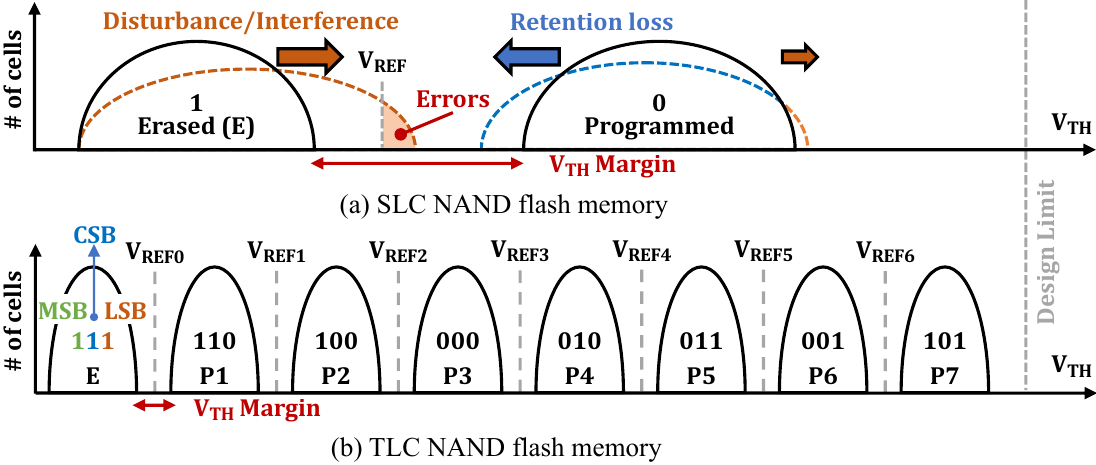}
     \caption{\vth distributions of NAND flash memory cells.}
     \label{fig:flash_reliability}
     \par\addvspace{-16pt}
\end{figure}

\head{NAND Flash Reliability}
NAND flash memory is error-prone due to a variety of error sources, such as read disturbance~\cite{ren-nvmsa-2023, cai-dsn-2015, ha-tcad-2016, zambelli-irps-17, liu-asplos-21}, program interference~\cite{park-dac-2016, kim-dac-2017, cai-hpca-2017, cai-iccd-2013}, and retention loss~\cite{luo-acm-2018, liu-target-2012, mizoguchi-imw-2017, yang-tcad-2025}.
\fig{\ref{fig:flash_reliability}} shows the \vth distribution of a WL, when programmed in (a) single-level cell (SLC) and (b) triple-level cell (TLC) modes to store one and three bits per cell, respectively.
Read and program operations unintentionally program other non-target WLs in the same block, slightly increasing the cells' \vth level (i.e.,  disturbance and interference).
A flash cell leaks its charge over time, which decreases its \vth level (i.e., retention loss).
If a cell's \vth level shifts beyond \vref, sensing the cell leads to a different value from the programmed originally, causing a bit error.
Two major factors significantly degrade the reliability of NAND flash memory.
First, the multi-level cell (MLC) technique drastically reduces the \vth margins to pack more \vth states within the limited voltage window.
Second, the high voltage applied to target cells during program and erase operations physically damage the cells, making it easier for the cells to gain or lose electrons more easily.

\head{Reliability Management}
There are four reliability-management techniques widely adopted to guarantee the integrity of stored data: \inum{i}~ECC, \inum{ii}~read retry, \inum{iii}~data refresh, and \inum{iv}~patrol read.
First, ECC detects and corrects raw bit errors in a read page~\cite{ldpc_decision, vahabzadeh-fms-2016, declercq-fms-2019}.
After reading the data from the flash memory, the flash controller detects and corrects the errors using the ECC parity that is stored in the out-of-bound (OOB) area of the same WL~\cite{ldpc_decision, vahabzadeh-fms-2016, declercq-fms-2019, ldpc, micron-flyer-2016}.
Second, if a read page's raw bit-error rate (RBER) exceeds the ECC capability, the SSD controller performs read retry that reads the page again with adjusted \vref levels until sufficiently lowering the page's RBER~\cite{park-asplos-2021, chun-hpca-2024}.
Third, to avoid potential data corruptions due to excessive RBER that even read retry cannot handle, the SSD controller proactively refreshes stored data.
For example, if a block's read count (i.e., the number of page reads to the block) exceeds a threshold (e.g., 400K~\cite{chun-cal-2025}), the SSD controller can eliminate transient errors stemming from read disturbance by copying the block's valid data to free pages, which is called \emph{read reclaim}.
Fourth, to proactively mitigate the overheads of ECC-decoding and read retry, some SSDs employ \emph{patrol reads}, which periodically read stored pages to \inum{i}~check their RBER and \inum{ii}~identify the near-optimal \vref levels for future reads~\cite{zhou-patrolread-2024, tanaka-patrolread-2016, alhussien-patrolread-2019}.

These reliability-management techniques enable modern SSDs to meet the strict unrecoverable bit-error ratio (UBER) requirements of $<$\(10^{-15}\)~\cite{jedec-jesd218}, despite the continuously degrading NAND flash reliability.
In practice, SSD controllers escalate to stronger and more expensive techniques as RBER increases.
For example, an SSD controller first performs hard-decision ECC decoding, which provides the lowest latency but also the weakest error-correction capability.
If hard-decision ECC decoding fails, the controller invokes more expensive recovery steps, such as read retry and soft-decision ECC decoding, and ultimately performs the data-refresh operations (e.g., read reclaim) when necessary. 
This progressive recovery process allows SSDs to successfully service nearly all I/O requests even under worst-case operating conditions (e.g., 1-year retention at 1.5K~PEC~\cite{cox-fms-2011, micron-pec}) at the cost of degraded I/O performance.
When an error remains uncorrectable even after the whole recovery process, the SSD reports it to the host as a \emph{read failure} rather than returning corrupted data, i.e., a successful read completion implies that no data corruption has been detected.

\section{Motivation}\label{sec:motivation}
Read disturbance has gained growing attention recently due to its increasing reliability impact.
Reading a page causes higher disturbance in modern NAND flash memory for two key reasons.
First, the advanced multi-level cell (MLC) techniques require more sensing steps for a page read, e.g., up to four sensing steps in quad-level cell (QLC) NAND flash memory, which significantly increases the read latency and thus exposes non-target cells to high \vpass for a longer time.
Second, a single read operation disturbs more non-target pages, as the block size in modern NAND flash memory rapidly increases (e.g., recent QLC NAND flash memory contains 2,816 pages per block~\cite{micron-QLC-2022}).

The increasing reliability impact of read disturbance, in turn, can significantly degrade the I/O performance of modern SSDs, aggravating reliability-management overheads.
As the RBER of a larger number of pages increases more rapidly, the SSD controller needs to perform read retry and read reclaim more frequently, which introduces significant performance overhead due to additional read and copy operations, respectively.
Although prior studies have proposed various techniques to mitigate read retry overhead~\cite{park-asplos-2021,chun-hpca-2024}, read reclaim's copy overhead rather severely increases in modern NAND flash memory due to the increased block size, i.e., read reclaim needs to copy a large number of pages in the target block.
Such increased overheads can affect application performance more seriously in modern computing systems, where a variety of data-intensive applications consistently read massive amounts of data from high-performance SSDs.

Nevertheless, no prior work has yet thoroughly investigated the system-level performance impact of read disturbance in practice. 
To our knowledge, only one recent study~\cite{liu-asplos-21} has demonstrated that read reclaim can degrade the I/O response time and lifetime of commodity SSDs, which motivates adaptively delaying read reclaim and relying more on other reliability-management techniques to mitigate performance degradation.
In the prior study, however, most evaluated SSDs are outdated in terms of storage interface (SATA) and NAND flash technology (2D or early 3D NAND flash memory)~\cite{sp550,su800,su630,barracuda,wdblue,hynixgold31,860evo,860qvo,tr200,bx500}, and only one workload (16-KiB random reads) is used.
With such outdated SSDs and limited workload, it is difficult to provide sufficient insights for a comprehensive understanding of the performance impact of read disturbance in modern SSDs that serve diverse I/O patterns with high-bandwidth interface (e.g., NVMe) and high-density (yet low-reliability) NAND flash memory.

\textbf{Our goal} is to provide new insights into the system-level performance impact of read disturbance in modern NAND flash-based storage systems, which can help develop better storage architecture and optimize system software (e.g., operating-systems I/O stack and SSD firmware) to better cope with ever-increasing I/O-performance demands from data-intensive applications.
To this end, we conduct a comprehensive experimental study that characterizes the performance of modern NVMe SSDs under different access patterns and operating conditions.

\section{Characterization Methodology}\label{sec:methodology}
This section presents our methodology for characterizing the impact of read disturbance on the system-level performance in modern NAND flash-based SSDs.
\subsection{Testing Infrastructure}
\label{ssec:test_infra}

\head{Tested SSDs}
\tbl{\ref{tab:ssd_spec}} summarizes the key characteristics of the 15 commodity SSDs from 10 vendors that we test~\cite{samsung980pro, samsungpm9a1, wdblacksn850pro, seagatefirecuda530, micron3400, kingstonfuryrenegade, hynixp41, crucialp3plus, micron2400, solidigmp41plus, hpfx900pro, solidigmp44pro, kioxiaxg8, crucialt500, hpfx700}. 
We characterize recent M.2 NVMe SSDs that employ advanced 3D TLC/QLC NAND flash memory with 96--232 stacked-WL layers.
To analyze the impact of SSD firmware, we test several SSDs that use the same NAND flash chips, i.e., chips from the same NAND manufacturer with the same number of WL layers and the same MLC technology.
All tested SSDs support PCIe Gen~4 interface (four lanes), offering high sequential-read bandwidth. 
For rapid characterization, we select SSD models that provide 500-GB or 512-GB storage capacity.
Even though process variation across NAND flash chips~\cite{shim-micro-2019,luo-pomacs-2018,wang-tecs-2017,chen-dac-2017,hung-jscc-2015,yen-hpca-2022,li-micro-2020,cai-date-2013,cai-dsn-2015,cho-asplos-2024} may lead to different performance behaviors even across SSDs of the same model, such within-model differences are not the focus of this work.
We expect any remaining differences to have limited impact on the system-level behaviors investigated in this study, given that SSD manufacturers generally aim to deliver consistent performance across devices of the same model (e.g., by using NAND flash chips with similar characteristics or wafer locations).
We leave a systematic investigation of within-model performance variation for future work.

\begingroup
\def\arraystretch{1}
\begin{table}[t]
  \caption{Specifications of 15 commodity SSDs tested.}
  \label{tab:ssd_spec}
  \centering
  \resizebox{\columnwidth}{!}{%
    \begin{tabular}{cccccccc}
      \toprule
      \multirow{2.5}{*}{\textbf{\makecell[c]{SSD ID}}} &
      \multicolumn{2}{c}{\textbf{Read Performance}} &
      \multicolumn{4}{c}{\textbf{Hardware Specification}} &
      \multirow{2.5}{*}{\textbf{\makecell[c]{Release Year}}} \\
      \cmidrule(lr){2-3}\cmidrule(lr){4-7}
      & \textbf{\makecell{Seq. \textbf{[MB/s]}}} & \textbf{\makecell{Rand. \textbf{[IOPS]}}} &
        \textbf{\makecell{NAND Mfr.}} &
        \textbf{\makecell{WL Layers/MLC}} &
        \textbf{\makecell{Capacity}} &
        \textbf{\makecell{Int. DRAM}} & \\[-0.5ex]
      \midrule
      \textbf{\ssdlist{A-samsung980}}     & 6,400 & 800K & \samsung  & 128/TLC & 500~GB & 512~MB & 2020 \\
      \textbf{\ssdlist{B-wd850}}          & 7,000 & 800K & \toshiba  &  96/TLC & 500~GB & 512~MB & 2020 \\
      \textbf{\ssdlist{C-solidigmP41}}    & 4,125 & 390K & \solidigm & 144/QLC & 512~GB & DRAM-less & 2022 \\
      \textbf{\ssdlist{D-samsungPM9A1}}   & 6,900 & 800K & \samsung  & 128/TLC & 512~GB & 512~MB & 2020 \\
      \textbf{\ssdlist{E-kingstonFR}}     & 7,300 & 450K & \micron   & 176/TLC & 500~GB & 512~MB & 2021 \\
      \textbf{\ssdlist{F-crucialP3}}      & 4,700 &  N/A & \micron   & 232/TLC & 500~GB & DRAM-less & 2022 \\
      \textbf{\ssdlist{G-hynixP41}}       & 7,000 & 960K & \hynix    & 176/TLC & 500~GB & 512~MB & 2022 \\
      \textbf{\ssdlist{H-kioxiaXG8}}      & 7,000 & 750K & \kioxia   & 112/TLC & 512~GB & 512~MB & 2022 \\
      \textbf{\ssdlist{I-micron2400}}     & 4,200 & 400K & \micron   & 176/QLC & 512~GB & DRAM-less & 2022 \\
      \textbf{\ssdlist{J-crucialT500}}    & 7,200 & 800K & \micron   & 232/TLC & 500~GB & 512~MB & 2023 \\
      \textbf{\ssdlist{K-seagateFC530}}   & 7,000 & 400K & \micron   & 176/TLC & 500~GB & 512~MB & 2021 \\
      \textbf{\ssdlist{L-hp900}}          & 7,000 & 540K & \micron   & 176/TLC & 512~GB & 512~MB & 2022 \\
      \textbf{\ssdlist{M-solidigmP44}}    & 7,000 & 960K & \hynix    & 176/TLC & 512~GB & 512~MB & 2022 \\
      \textbf{\ssdlist{N-hp700}}          & 6,300 & 567K & \ymtc     & 232/QLC & 512~GB & DRAM-less & 2023 \\
      \textbf{\ssdlist{O-micron3400}}     & 6,600 & 360K & \micron   & 176/TLC & 512~GB & 1~GB & 2021 \\
      \bottomrule
    \end{tabular}
    }
    \par
\end{table}
\endgroup

\head{SSD Testing Setup}
We set up our test environment using eight Linux machines, each containing an Intel i9-14900K CPU~\cite{intel149000k}, four 16-GB DDR4-3200~MHz DRAM modules, and a 4-TB Samsung 990 Pro SSD~\cite{samsung-990pro}.
We run FIO benchmark tool~\cite{fio} to issue I/O requests to the tested SSDs while collecting log data for every 100~ms using the main storage device.\footnote{We set the logging interval to 100~ms to reduce the log file size, which we verify is sufficient to capture all performance patterns observed at shorter intervals (e.g., 5~ms).}
While running FIO, we \inum{i}~enable a direct I/O option to bypass the host memory buffer when accessing tested SSDs and \inum{ii}~use the libaio I/O engine to saturate SSD performance.
To avoid potential interference, we evaluate only one SSD at a time on each machine.
We check S.M.A.R.T. logs~\cite{nvmespec, smartctl, msecli} to obtain internal information of tested SSDs, such as their remaining lifetime and operating temperature.
Each SSD is installed with a heat sink to minimize thermal throttling, allowing the SSD to operate at a stable temperature range of \degreec{40--75}, below \emph{Warning Temperature Threshold} (\degreec{85}) specified by the S.M.A.R.T. attribute.
We do not observe any temperature-induced throttling when using a heat sink, which can otherwise occur when running experiments without one.

\subsection{Testing Methodology}
\label{ssec:test_method}
\head{SSD Preconditioning}
We precondition all SSDs for each read-disturbance test in three steps to avoid potential distortions caused by any factor other than read disturbance. 
First, we issue TRIM commands to the entire logical block address (LBA) range to minimize possible variations in data layout and garbage collection.
Second, we fill up 90\% of SSD logical capacity, a host-visible LBA space which excludes over-provisioned area and metadata, by sequentially writing 450~GB of data starting from LBA \texttt{0}.
Doing so ensures that the read data has been programmed in TLC/QLC blocks rather than in the SLC buffer~\cite{yoo-hotstorage-2020, tanpairoj-slccache-2022}. 
Third, we give a 30-minute idle time before each test, which provides sufficient time for SSDs to perform internal tasks such as garbage collection and SLC-buffer flush as they need.

\head{Read-Disturbance Test}
We analyze the performance impact of read disturbance in the 15 SSDs while varying access patterns and SSD-lifetime stages.
We accumulate read disturbance to SSDs under sequential- and random-read workloads generated from FIO with the configurations widely used for SSD-performance specification~\cite{phison-random-read}: \inum{i}~queue depths (QD) of 32 and 128, \inum{ii}~I/O sizes of 1~MiB and 4~KiB, and \inum{iii}~thread counts of 1 and 16, respectively.
We expect this configuration to be sufficient to fully saturate the NAND flash chips, thus minimizing performance impacts from plane-level load imbalance.
We perform tests while varying the consecutive LBA range to be read, referred to as \emph{read range}, from 64~MB to 50~GB;
we read the middle of the 450-GB preconditioned data (e.g.,~50-GB reads starting at a 200-GB offset) to avoid reading data in the SLC buffer.
This is because we observe that some tested SSDs retain either the oldest or the newest data in the SLC buffer even after long idle times.
Reading such data yields artificially high performance, which hinders characterizations of common and worst-case scenarios.
All read requests complete successfully without errors in all tests, i.e., no data corruption occurs, as explained in \sect{\ref{ssec:bg_nand_rel}}.

We evaluate SSDs under three lifetime stages, early, middle, and late, each corresponding to 0\%, 50\%, and 80\% of their lifetime usage, respectively.
This is motivated by prior findings that read disturbance becomes more severe with higher program/erase (P/E) cycles~\cite{liu-tos-2022, ren-nvmsa-2023}. 
To \emph{uniformly} increase P/E-cycle counts (PEC), we repeat sequential writes to the entire LBA range of the SSDs. 
We observe that some SSDs occasionally produce variations in test results.
We hypothesize that these are caused by different data layouts across tests, which is challenging to control.
For such SSDs, we repeat the tests five times or more and use the most dominant result.

\section{Characterization Results}\label{sec:characterization_results}
We present our characterization results on the system-level performance impact of read disturbance in modern SSDs, introducing eleven new observations and five key takeaways.

\subsection{Impact of Access Patterns}\label{ssec:access_pattern_test}
\begin{figure}[h]
     \centering
     \includegraphics[width=\linewidth]{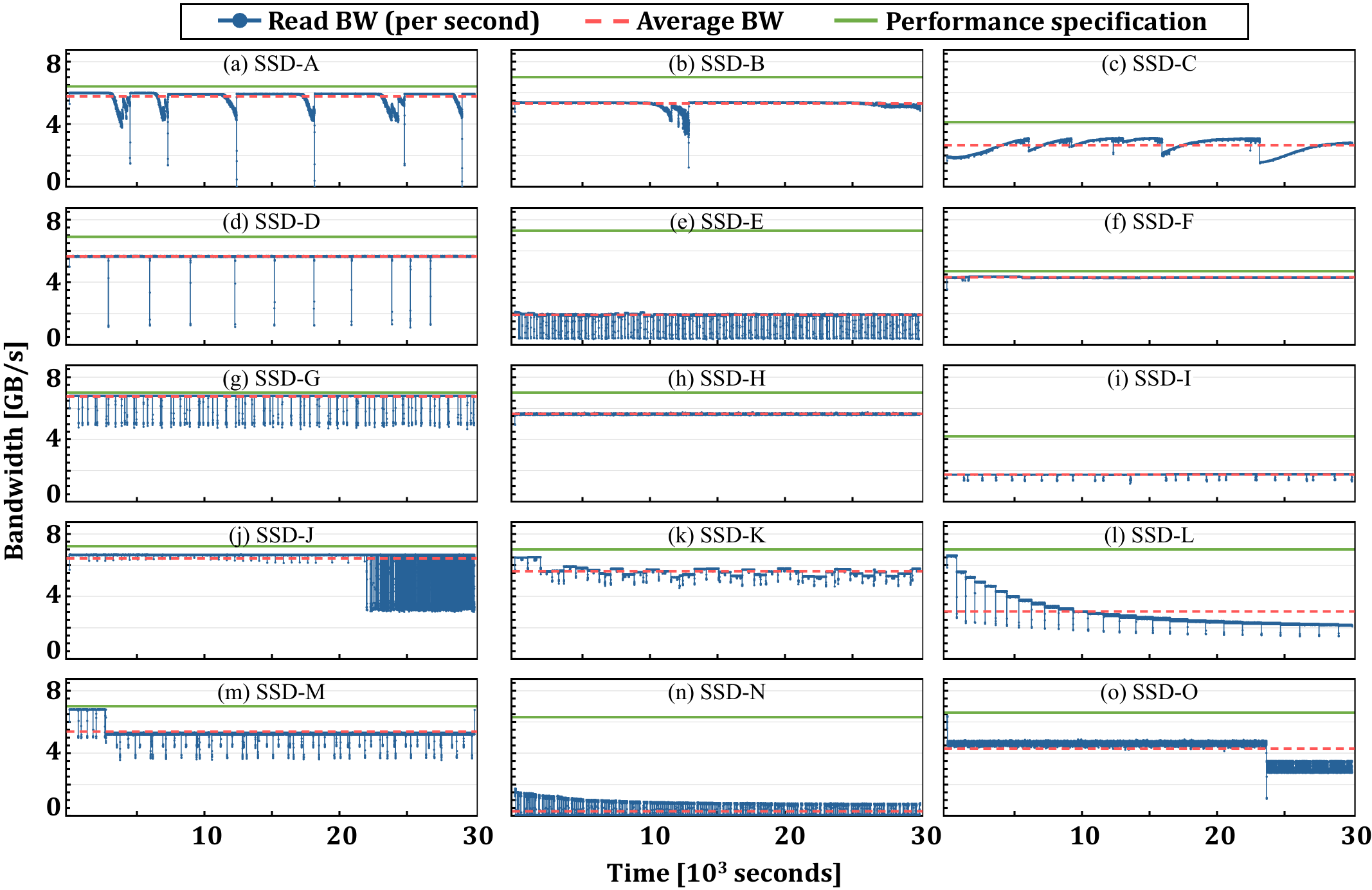}
     \caption{Sequential-read test results with 1~GB read range.}
     \label{fig:seq_1GB}
     \par\addvspace{-10pt}
\end{figure}

We first analyze how read disturbance affects SSD performance under sequential and random reads.
\head{Sequential-Read Pattern}
\fig{\ref{fig:seq_1GB}} shows the performance of the 15 SSDs during the sequential-read test for 30,000 seconds with a read range of 1 GB at early lifetime stages.
Each data point indicates the amount of data read during one second.
We also plot \inum{i}~the sequential-read performance specified by the manufacturer and \inum{ii}~the average bandwidth actually measured during the test.

We make three key observations from \fig{\ref{fig:seq_1GB}}.

\observation{obs:seq_specification}{A majority of the tested SSDs fail to match their performance specifications.}
All SSDs except for \ssdlist{F-crucialP3} and \ssdlist{G-hynixP41} exhibit significantly lower performance compared to the specification by 36.74\% on average (up to 95.5\% in \ssdlist{N-hp700}).
\ssdlist{E-kingstonFR, I-micron2400, L-hp900, N-hp700} fail to achieve half the specified performance even at the beginning of the tests.
The results suggest that \inum{i}~the specified performance could be measured in the best-case scenario, e.g., when reading data stored in the SLC buffer~\cite{tanpairoj-slccache-2022, yoo-hotstorage-2020}, and \inum{ii}~an SSD may not meet its performance specification in practice even under a sequential-read workload (which is considered best for SSD performance~\cite{ghiasi-asplos-2022, alizadeh-arxiv-2023, wang-hpca-2024, ghiasi-isca-2024}).

\observation{obs:seq_fluctuation}{Most tested SSDs exhibit significant performance fluctuations consistently.}
\tbl{\ref{tab:seq_drop}} summarizes each SSD's performance-drop patterns, where the drop duration and period represent the time required for an SSD to recover near-peak bandwidth and the time until the next performance drop, respectively, since a drop occurs.
All tested SSDs except for \ssdlist{F-crucialP3} and \ssdlist{H-kioxiaXG8} exhibit a significant gap between the peak and minimum bandwidth values.
In particular, \ssdlist{A-samsung980}'s read bandwidth periodically drops to \mytilde1~MB/s, which means that it can hardly service user reads for one second, significantly increasing the read tail latency.
The 1st-percentile (P1) bandwidth is close to the minimum bandwidth ($<$20\% differences) in half of the SSDs (\ssdlist{C-solidigmP41, E-kingstonFR, G-hynixP41, I-micron2400, J-crucialT500, K-seagateFC530, M-solidigmP44, N-hp700}), and the 10th-percentile (P10) bandwidth is lower than the peak bandwidth by more than 30\% in \ssdlist{C-solidigmP41, L-hp900, N-hp700, O-micron3400}.
The results highlight the consistent performance impact of read disturbance in modern SSDs.
Such performance fluctuations, in turn, considerably degrade the average bandwidth in most tested SSDs compared to their peak bandwidth by 22.1\% on average (up to 83.8\% in \ssdlist{N-hp700}).

\begingroup
\setlength{\tabcolsep}{7pt}
\def\arraystretch{0.9}
\begin{table}[h]
  \caption{Performance-drop patterns during the sequential read test with a read range of 1~GB.}
  \label{tab:seq_drop}
  \centering
  \resizebox{\columnwidth}{!}{%
    \begin{tabular}{cccccccc}
      \toprule
      \multirow{2.5}{*}{\textbf{\makecell{SSD \\ ID}}} &
      \multicolumn{5}{c}{\textbf{Read Bandwidth (Normalized to Peak)~\textbf{[MB/s]}}} &
      \multicolumn{2}{c}{\textbf{Drop Timing [seconds]}} \\
      \cmidrule(lr){2-6}\cmidrule(lr){7-8}
          & \textbf{Peak} & \textbf{Avg.} & \textbf{Min.} & \textbf{P1.} & \textbf{P10.} &
            \textbf{Duration} & \textbf{\makecell{\minorm{Avg. Period}}} \\
      \toprule
      \textbf{\ssdlist{A-samsung980}}  & 6,026 & 5,768 (95.7\%) & 1 ($<$0.1\%)     & 4,344 (72.1\%) & 5,145 (85.4\%) & 775--1.7K    & 3.9K \\
      \textbf{\ssdlist{B-wd850}}       & 5,423 & 5,313 (98\%)   & 1,231 (22.7\%)   & 4,396 (81.1\%) & 5,153 (95\%)   & 2.2K        & $>$10K \\
      \textbf{\ssdlist{C-solidigmP41}} & 3,144 & 2,650 (84.3\%) & 1,504 (47.8\%)   & 1,586 (50.4\%) & 1,929 (61.4\%) & 2.3K--3.5K   & 2.7K \\
      \midrule
      \textbf{\ssdlist{D-samsungPM9A1}}& 5,695 & 5,646 (99.1\%) & 1,099 (19.3\%)   & 5,621 (98.7\%) & 5,636 (99\%)   & 2--4         & 2.7K \\
      \textbf{\ssdlist{E-kingstonFR}}  & 2,118 & 1,914 (90.4\%) &   346 (16.3\%)   &   432 (20.4\%) & 1,883 (88.9\%) & 1--2         & 156 \\
      \textbf{\ssdlist{F-crucialP3}}   & 4,371 & 4,307 (98.5\%) & 4,102 (93.8\%)   & 4,286 (98.1\%) & 4,294 (98.2\%) & N/A$^1$     & N/A$^1$ \\
      \textbf{\ssdlist{G-hynixP41}}    & 6,818 & 6,767 (99.3\%) & 4,674 (68.6\%)   & 5,073 (74.4\%) & 6,794 (99.6\%) & 5--7         & 331 \\
      \textbf{\ssdlist{H-kioxiaXG8}}   & 5,761 & 5,649 (98.1\%) & 5,546 (96.3\%)   & 5,597 (97.2\%) & 5,622 (97.6\%) & N/A$^1$     & N/A$^1$ \\
      \textbf{\ssdlist{I-micron2400}}  & 1,786 & 1,749 (97.9\%) & 1,172 (65.6\%)   & 1,389 (77.8\%) & 1,739 (97.4\%) & 15--65       & 1.0K \\
      \textbf{\ssdlist{J-crucialT500}} & 6,692 & 6,427 (96\%)   & 3,011 (45\%)     & 3,200 (47.8\%) & 6,643 (99.3\%) & 1--5         & 13 \\
      \midrule
      \textbf{\ssdlist{K-seagateFC530}}& 6,531 & 5,611 (85.9\%) & 4,529 (69.3\%)   & 4,988 (76.4\%) & 5,293 (81\%)   & 12--21       & 672 \\
      \textbf{\ssdlist{L-hp900}}       & 6,638 & 3,039 (45.8\%) & 1,456 (21.9\%)   & 2,143 (32.3\%) & 2,197 (33.1\%) & 3--4         & 1.1K \\
      \textbf{\ssdlist{M-solidigmP44}} & 6,813 & 5,388 (79.1\%) & 3,573 (52.4\%)   & 4,479 (65.7\%) & 5,207 (76.4\%) & 5--7         & 420 \\
      \textbf{\ssdlist{N-hp700}}       & 1,741 &   282 (16.2\%) &    58 (3.3\%)    &    59 (3.4\%)  &    59 (3.4\%)  & 1--243       & 37 \\
      \midrule
      \textbf{\ssdlist{O-micron3400}}  & 6,363 & 4,299 (67.6\%) & 1,084 (17\%)     & 2,776 (43.6\%) & 2,821 (44.3\%) & N/A$^2$     & N/A$^2$ \\
      \bottomrule
    \end{tabular}
    }
    \par
    \vspace{2pt}
    \parbox{\columnwidth}{\footnotesize \textbf{N/A:} $^1$no significant drops or $^2$too-short drop duration and period. Values in parentheses are percentages of the peak bandwidth for each SSD.}
    \par\addvspace{-5pt}
\end{table}
\endgroup

We observe that \ssdlist{F-crucialP3} and \ssdlist{H-kioxiaXG8} provide stable performance during the entire test.
We hypothesize that those SSDs can hide the performance impact of read disturbance by efficiently leveraging their over-provisioned internal parallelism.
Most tested SSDs, including \ssdlist{F-crucialP3} and \ssdlist{H-kioxiaXG8} contain NAND flash chips more than enough to achieve their specified sequential-read performance.
This means that they can perform read-disturbance management tasks using such extra resources without affecting the average bandwidth.
Note that, however, when we extend the test time to 1,000,000~seconds, performance drops also occur in \ssdlist{F-crucialP3} and \ssdlist{H-kioxiaXG8} at around 45,000- and 83,000-second periods, respectively, which suggests that it is challenging to completely eliminate the performance impact of read disturbance in modern SSDs.

\observation{obs:drop_pattern}{There exist distinct performance-drop patterns observed across the tested SSDs.}
We categorize the tested SSDs into two groups based on whether they attempt to maintain near-peak performance (\textbf{\typeA}) or not (\textbf{\typeB}) using their performance-drop patterns as a proxy for SSD-internal behavior, since analyzing exact SSD-internal behavior is practically infeasible.
We hypothesize that \typeA SSDs (\ssdlist{A-samsung980} to \ssdlist{J-crucialT500}) actively perform read-disturbance management tasks, such as read reclaim and patrol reads, which cause steep yet short performance drops but prevent long-term slowdowns.
In contrast, \typeB SSDs (\ssdlist{K-seagateFC530} to \ssdlist{O-micron3400}) appear to defer read reclaim as much as possible while relying on other reliability-management techniques, e.g., read retry~\cite{park-asplos-2021} and strong ECC~\cite{chun-hpca-2024}, to avoid significant increases in read tail latency.
We validate our hypothesis based on the SSD-internal statistics that can be obtained for some SSDs\footnote{Most tested SSDs do not disclose their internal statistics.}
from the vendor-specific S.M.A.R.T. attribute.
Once performance fluctuation occurs in \ssdlist{E-kingstonFR} and \ssdlist{J-crucialT500} (\typeA), their \emph{device lifetime percentage} and \emph{internal flash write} values increase by up to 1\% and 516~GB, respectively, even though no user write has been issued, which strongly suggests that the performance fluctuation is likely to be caused by read reclaim.
In contrast, such attributes never increase in \ssdlist{K-seagateFC530} to \ssdlist{O-micron3400} (\typeB) during the entire test.
Note that, however, detailed performance-drop patterns significantly vary even across SSDs using the same NAND flash chips, e.g., \ssdlist{E-kingstonFR, K-seagateFC530, L-hp900, O-micron3400}.

\begin{figure}[h] 
     \centering
     \includegraphics[width=0.95\linewidth]{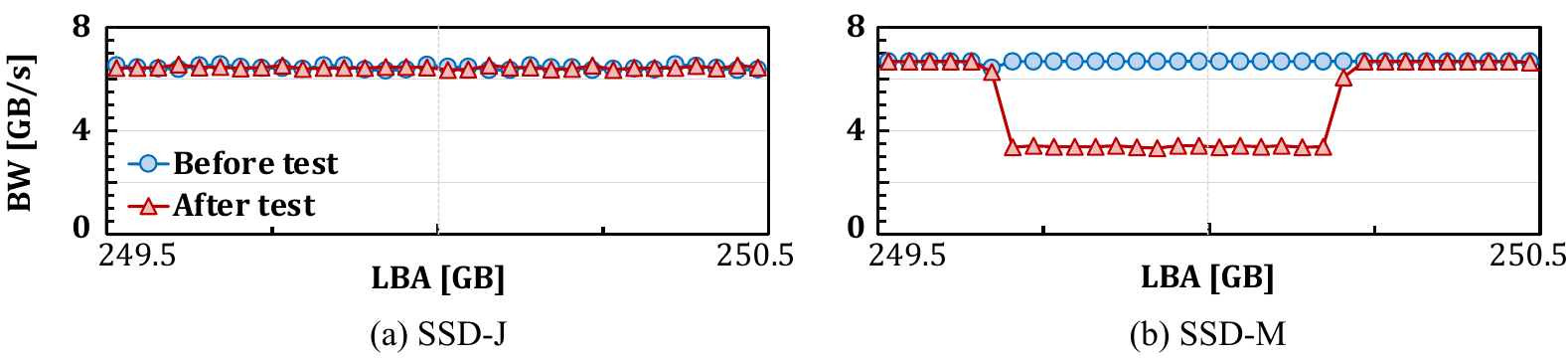}
     \par\addvspace{-5pt}
     \caption{Read bandwidth breakdown before and after test.}
     \label{fig:seq_perf_breakdown}
\end{figure}

To identify the root cause of the peak-performance degradation in \typeB SSDs, we further analyze the performance impact of read disturbance more precisely.
To this end, before and after a sequential-read test with a 1-GB read range, we profile the sequential-read bandwidth of each 32-MB region within the read range.
\fig{\ref{fig:seq_perf_breakdown}} compares the read bandwidth across the 32-MB regions before and after the sequential-read test in (a)~\ssdlist{J-crucialT500} (\typeA) and (b)~\ssdlist{M-solidigmP44} (\typeB) at early lifetime stages.

\observation{obs:weak_WLs}{Certain parts of the read range exhibit consistently degraded performance in the \typeB SSDs, whereas the \typeA SSDs maintain their peak performance across the entire read range after the sequential-read test.}

Half of the 32-MB regions in \ssdlist{M-solidigmP44} incur 49.05\% bandwidth degradation on average compared to other regions within the 1-GB read range.
We observe the same trends for all \typeB SSDs, which also supports our hypothesis that they rely on read retry or strong ECC rather than read reclaim or patrol reads.
Prior work has demonstrated that some WLs are more susceptible to error sources than other WLs~\cite{chun-cal-2025, shim-micro-2019, xiong-tos-2018}.
As the RBER of such \emph{weaker} WLs increases more rapidly compared to other WLs with the same number of page reads to the block, read disturbance can lead them to incur read retry and long ECC decoding more easily~\cite{chun-hpca-2024, dong-tocs-2013, zhang-sips-2015}, increasing the read latency.
As read disturbance accumulates, the fraction of WLs with long read latencies increases, which continuously degrades read bandwidth.
Read reclaim and patrol reads can avoid such bandwidth degradation by directly reducing the RBER of read data as shown in \fig{\ref{fig:seq_perf_breakdown}(a).}

\take{Read disturbance can significantly degrade SSD performance even under sequential reads, but its impact varies across SSDs (even those using the same NAND flash chips) depending on read-disturbance management.}

\begin{figure}[h]
     \centering
     \vspace{-1em}
     \includegraphics[width=\linewidth]{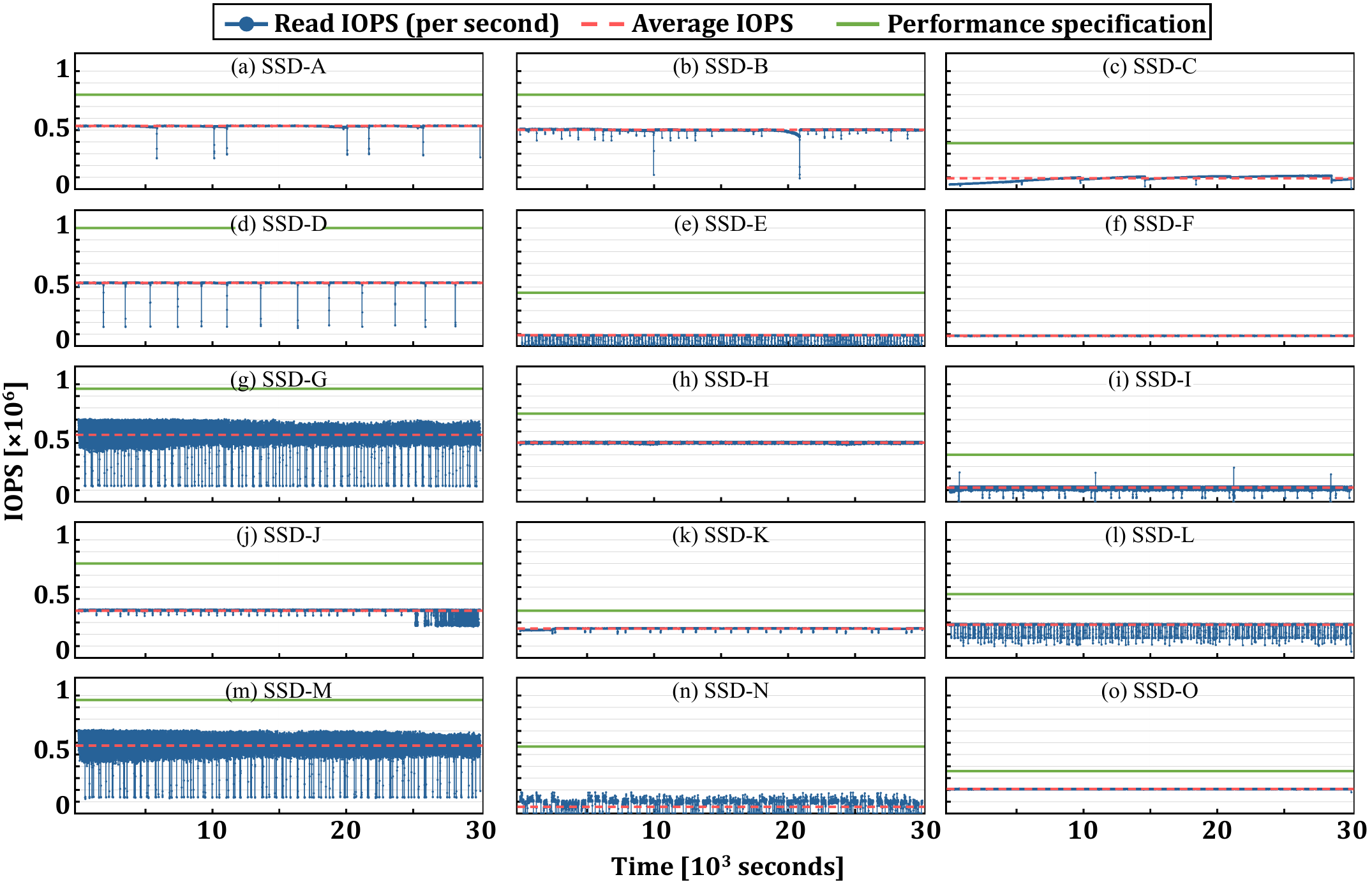}
     \caption{Random-read test results with 1 GB read range.}
     \label{fig:rand_1GB}
\end{figure}

\head{Random-Read Pattern}
\fig{\ref{fig:rand_1GB}} shows the performance of the 15 SSDs during the random-read test for 30,000 seconds with a read range of 1~GB at early lifetime stages.
Each data point indicates the number of read requests serviced during one second (IOPS).
We also plot the random-read performance specified by the manufacturer and average IOPS measured during the entire test.

We make two key observations from \fig{\ref{fig:rand_1GB}}.

\observation{obs:rand_specification}{All tested SSDs exhibit significantly lower random-read performance than the specification, showing even larger performance gaps than in the sequential-read test.}
The measured IOPS is far lower than the specification by 51.93\% on average across all tested SSDs,\footnote{We exclude \ssdlist{F-crucialP3} due to its lack of random-read performance specification publicly accessible. Note that \ssdlist{F-crucialP3} merely provides similar or lower IOPS compared to other DRAM-less QLC SSDs (\ssdlist{C-solidigmP41} and \ssdlist{I-micron2400}), which suggests that the actual performance gap could be even larger with \ssdlist{F-crucialP3}.}
which is much more severe than in the sequential-read test (36.74\%).
The result further supports our hypothesis in \cref{obs:seq_specification} that most SSD manufacturers determine the performance specification in the best-case scenario.
It is well known that SSDs \emph{cannot} deal with small random reads as efficiently as sequential reads due to two reasons. 
First, the minimum I/O unit of NAND flash memory is a page (typically 16~KiB), so reading sub-page data \emph{cannot} increase throughput (IOPS), merely decreasing bandwidth.
Second, the independent-plane read has been recently introduced and may not yet be adopted commonly in modern SSDs; until recently, most NAND flash chips could perform multi-plane operations \emph{only when} target pages have the same block or page offsets~\cite{onfispec}, which is unlikely to hold under random reads, thereby significantly under-utilizing SSD-internal parallelism compared to sequential reads.

\observation{obs:rand_less_impact}{For a majority of the tested SSDs, the performance impact of read disturbance is less significant under random reads compared to under sequential reads.}
We observe two key differences in the performance-drop patterns between the random- and sequential-read tests.
First, no \typeB SSDs (\ssdlist{K-seagateFC530} to \ssdlist{O-micron3400}) exhibit long-term peak-performance degradation during the random-read test, unlike during the sequential-read test.
Second, a majority of the tested SSDs (\ssdlist{A-samsung980, B-wd850, D-samsungPM9A1, E-kingstonFR, F-crucialP3, H-kioxiaXG8, J-crucialT500, K-seagateFC530, L-hp900, O-micron3400}) incur less significant performance fluctuation during the random-read test, showing smaller gaps between their peak and minimum performance (46.5\% on average) than those in the sequential-read test (59.8\%).
The results align with our initial expectation; 
even a slight decrease in SSD-internal parallelism caused by read disturbance can significantly affect the peak performance for sequential-read workloads compared to random-read workloads (where the available parallelism is inherently limited).
This highlights the importance of analyzing the performance impact of read disturbance with various workloads, suggesting that the performance impact of read disturbance demonstrated in prior work~\cite{liu-asplos-21} (which has used only random-read workloads) might be underrated.

Read disturbance-induced performance fluctuations can also be more severe in a few SSDs under random-read workloads due to two reasons.
First, we hypothesize that high address-translation overhead under random-read pattern can exacerbate performance fluctuations in DRAM-less SSDs (\ssdlist{C-solidigmP41, I-micron2400, N-hp700}).
DRAM-less SSDs store most L2P mappings in NAND flash memory~\cite{zhang-dramless-2020, gupta-asplos-2009}, which frequently incurs additional NAND-flash access for address translation\minorm{, and, in turn, can amplify the overheads of internal reliability management that require address-mapping information.}
Under sequential reads, high spatial locality across accesses to L2P mappings minimizes the address-translation overhead, but it would significantly increase under random reads.
Second, we assume that \ssdlist{G-hynixP41} and \ssdlist{M-solidigmP44} heavily rely on locality-sensitive information for reliability management.
For example, several prior works have proposed to reuse near-optimal \vref levels of recently read WLs for future reads~\cite{shim-micro-2019, cai-iccd-2013, cai-dsn-2015, luo-jsac-2016, luo-hpca-2018, luo-sigmetrics-2018, nie-dac-2020}.
Doing so can potentially mitigate read-retry overheads for neighboring WLs of the recently read ones that would likely have similar reliability characteristics.
Unfortunately, its effectiveness may degrade significantly under random reads due to the limited space for tracking recently used \vref levels, thereby requiring more aggressive read-disturbance management.

\take{The performance impact of read disturbance is generally worse under sequential-read workloads (which are considered best for SSD performance in a large body of work~\cite{ghiasi-asplos-2022, alizadeh-arxiv-2023, wang-hpca-2024, ghiasi-isca-2024}) than under random-read workloads.}

\noindent 
From our later analysis, we focus on sequential-read pattern rather than random-read pattern, since read disturbance has a more significant impact on sequential-read performance.

\subsection{Impact of Read Ranges}\label{ssec:read_range_test} 
\begin{figure}[h] 
     \centering
     \includegraphics[width=\linewidth]{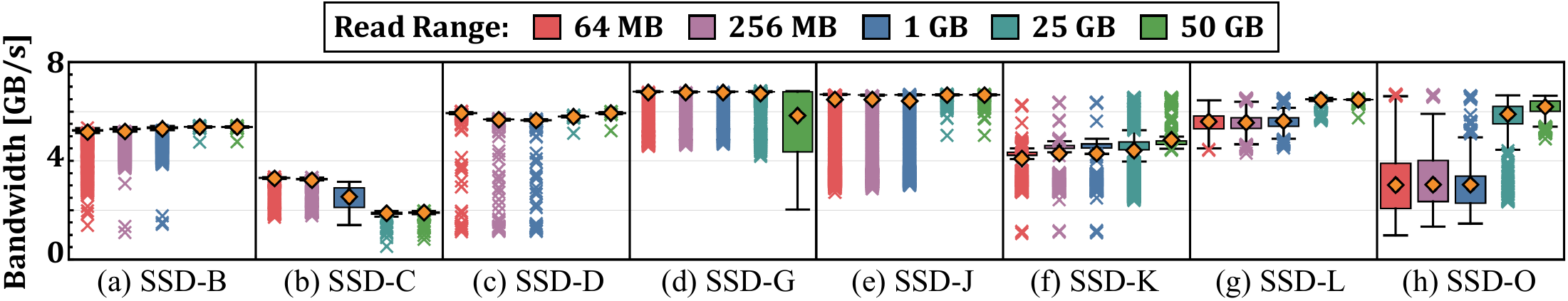}
     \par\addvspace{-5pt}
     \caption{Read-bandwidth distribution with different read ranges.}
     \label{fig:seq_readrange}
     \par\addvspace{-5pt}
\end{figure}

To better understand the read-disturbance management in modern SSDs, we perform the sequential-read tests while varying the read range from 64~MB to 50~GB, which would change the number of WLs and blocks repeatedly read during the test.
We assume that all tested SSDs evenly distribute incoming data across planes, which is common practice to fully leverage SSD-internal parallelism~\cite{chen-hpca-2011, jung-hpca-2014, kim-fast-2017}.
If our assumption holds, narrowing the read range to 64~MB or 256~MB limits the number of WLs repeatedly read within a block for all tested SSDs, whereas expanding the range to 25~GB or 50~GB leads multiple blocks to be disturbed during our sequential-read tests.
\fig{\ref{fig:seq_readrange}} compares the per-second bandwidth of eight SSDs (\ssdlist{B-wd850, C-solidigmP41, D-samsungPM9A1, G-hynixP41, J-crucialT500, K-seagateFC530, L-hp900, O-micron3400}) at early lifetime stage using box-and-whisker plots.
We select the eight SSDs that represent each SSD group with the distinct characteristics and exhibit acceptable read-disturbance management overheads.
We also plot the average bandwidth during the test (\yellowdiamond) and outliers ($\times$) in \fig{\ref{fig:seq_readrange}}.

We make three key observations from \fig{\ref{fig:seq_readrange}}.

\observation{obs:seq_small_similar}{Most tested SSDs show similar performance for read ranges from 64~MB to 1~GB.}
Except for \ssdlist{B-wd850} and \ssdlist{C-solidigmP41} (which will be discussed later in \cref{obs:seq_readrange_exception}), all SSDs exhibit largely consistent average bandwidth with variations smaller than 5.08\% across the 64-MB, 256-MB, and 1-GB read ranges.
In particular, the performance drop timings of \ssdlist{G-hynixP41, K-seagateFC530, O-micron3400} also hardly change with the read range (not directly shown in \fig{\ref{fig:seq_readrange}}).
The results strongly suggest that these SSDs perform read-disturbance management at the \emph{block} level, e.g., reclaiming a block when the block's read count exceeds a threshold~\cite{camp-readreclaim-2010, mataya-readreclaim-2014, han-vlsi-2023}.

\observation{obs:seq_large_stable}{Most tested SSDs better perform with larger read ranges, e.g., 25~GB and 50~GB.}
Except for \ssdlist{C-solidigmP41} and \ssdlist{G-hynixP41} (discussed later in \cref{obs:seq_readrange_exception}), all SSDs exhibit higher average bandwidth (\ssdlist{B-wd850, J-crucialT500, K-seagateFC530, L-hp900, O-micron3400}) and/or less performance fluctuations (\ssdlist{B-wd850, D-samsungPM9A1, J-crucialT500, K-seagateFC530, L-hp900, O-micron3400}) with the 25-GB and 50-GB read ranges compared to the small ranges.
This result can be readily explained, as distributing read requests across a larger range would reduce per-block read disturbance over the same time window, postponing read disturbance-induced performance degradation.
Note that, however, when we extend the test time up to 3,000,000~seconds, all SSDs eventually incur performance drops even more significant than with the small read ranges, as they need to simultaneously perform read-disturbance management tasks for more blocks, which clearly shows that read disturbance and its management are inevitable in modern SSDs.

\begin{figure*}[h]
     \centering
     \par\addvspace{-5pt}
     \includegraphics[width=\linewidth]{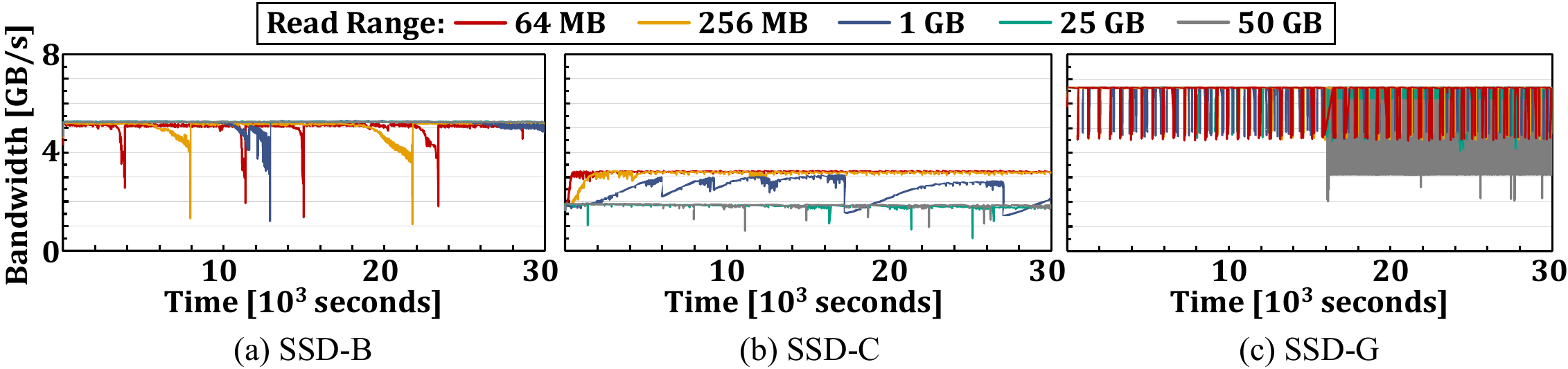}
     \caption{Comparisons of sequential-read bandwidth across different read ranges in outlier SSDs.}
     \par\addvspace{-5pt}
     \label{fig:seq_readrange_comp}
\end{figure*}

\observation{obs:seq_readrange_exception}{A few SSDs exhibit distinct behaviors across the read ranges from other SSDs.}
Unlike other SSDs, \ssdlist{B-wd850} and \ssdlist{C-solidigmP41} exhibit varying performance across the small (64-MB, 256-MB, and 1-GB) read ranges (opposite to \cref{obs:seq_small_similar}), and the average bandwidth of \ssdlist{C-solidigmP41} and \ssdlist{G-hynixP41} significantly degrades with the larger read ranges (opposite to \cref{obs:seq_large_stable}).
We further analyze these \emph{outlier} SSDs with more comprehensive comparisons of performance variations depending on the read range.
\fig{\ref{fig:seq_readrange_comp}} shows the read bandwidth of (a) \ssdlist{B-wd850}, (b) \ssdlist{C-solidigmP41}, and (c) \ssdlist{G-hynixP41}, during 30,000-second sequential-read tests while varying read ranges from 64 MB to 50 GB.

We observe that the three outlier SSDs operate significantly differently even compared to each other.
First, \ssdlist{B-wd850} incurs more frequent performance drops with a smaller read range: four, two, and one drop(s) with 64-MB, 256-MB, and 1-GB read ranges during the test, respectively.
The result implies that \ssdlist{B-wd850} is aware of the varying reliability impact of read disturbance within a block; prior studies have demonstrated that reading a WL disturbs its adjacent WLs more significantly compared to other WLs~\cite{chun-cal-2025, xiong-tos-2018, ren-nvmsa-2023}.
Repeated reads to only part of each block increase the RBER of the read WLs more rapidly compared to the others, thereby requiring more frequent management tasks for the WLs.
We hypothesize that other SSDs perform read-disturbance management \emph{more conservatively}, e.g., reclaiming an entire block even when some WLs can actually endure more read disturbance, which can minimize performance variations across the small read ranges.

Second, \ssdlist{C-solidigmP41} performs better with smaller read ranges, stably providing high bandwidth.
In fact, \ssdlist{C-solidigmP41} exhibits \emph{unique} performance-drop patterns during the 1-GB sequential-read test; its bandwidth gradually \emph{increases from the beginning} of the test until a drop occurs, whereas all other SSDs provide their peak bandwidth initially.
We hypothesize that \ssdlist{C-solidigmP41} progressively migrates frequently read pages to MLC/TLC regions, allowing the pages to provide higher read performance and endure more read disturbance than QLC pages. 
This would be effective for small read ranges, but it can cause frequent garbage collections due to migration, providing diminishing returns with larger read ranges as shown in \fig{\ref{fig:seq_readrange_comp}(b)}.

Third, like most other SSDs, \ssdlist{G-hynixP41} incurs less frequent performance drops with the 25-GB and 50-GB read ranges compared to the smaller ranges, \emph{but only during the first half} of the test; its performance fluctuation becomes more severe with the large ranges as the test continues.
In particular, with the 50-GB read range, significant and periodic performance drops continue after around 16,000 seconds until the end of the test, where the read bandwidth fluctuates from 6.8~GB/s (sustained for 4--5~seconds) to 3.2~GB/s (for 5--7~seconds).
During each drop period (i.e.,~9--12~seconds), \ssdlist{G-hynixP41} reads almost the same amount of data as the read range (i.e.,~50~GB), and we confirm that specific regions consistently exhibit degraded bandwidth, similarly to \cref{obs:weak_WLs}.
We hypothesize that \ssdlist{G-hynixP41} periodically performs patrol reads to track the near-optimal \vref levels of stored pages to avoid read retry for future reads.
Unfortunately, it is challenging to store the near-optimal \vref levels for every page due to the memory constraints, which limits the effectiveness of patrol reads when a large number of pages are continuously read.

\take{Even though various read-disturbance management techniques are employed across tested SSDs, there is no ultimate technique that operates optimally in all situations.}

\subsection{Impact of PEC}\label{ssec:pec_test} 
\begin{figure}[h]
     \centering
     \par\addvspace{-7pt}
     \includegraphics[width=0.8\linewidth]{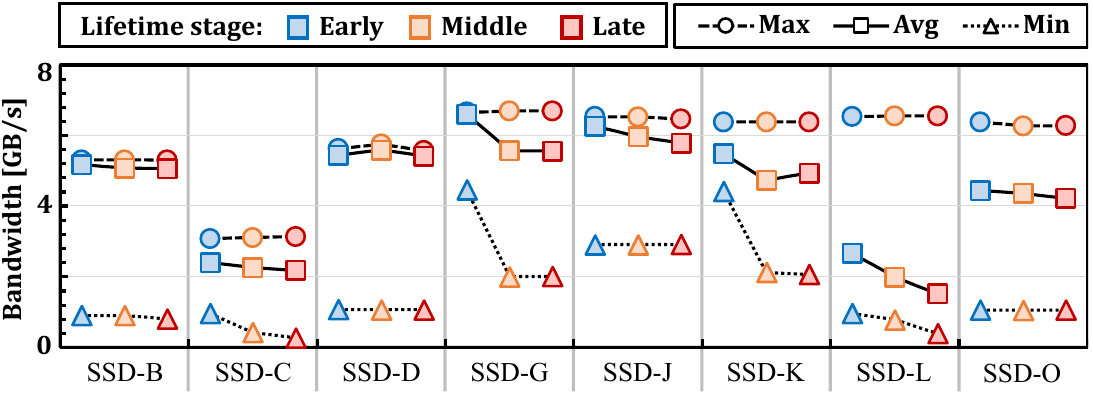}
     \caption{Read bandwidth at different device lifetime stages.}
     \label{fig:seq_pec}
     \par\addvspace{-10pt}
\end{figure}

We analyze how the performance impact of read disturbance changes depending on the SSD-lifetime stage, i.e., the average PEC across blocks.
To this end, we repeat the sequential-read tests with a 1-GB read range for eight SSDs at their middle ($>$50\%) and late ($>$80\%) lifetime stages.
\fig{\ref{fig:seq_pec}} compares the minimum, average, and maximum read-bandwidth values under the three lifetime stages (early: $<$20\% of SSD lifetime).

\observation{obs:seq_pec}{Read bandwidth generally decreases at later lifetime stages in all tested SSDs, but the degree of performance degradation varies across the SSDs.}
At the middle/late lifetime stages, most tested SSDs exhibit lower \inum{i}~average read bandwidth (by 11.2\%/15.5\% on average across \ssdlist{C-solidigmP41, G-hynixP41, J-crucialT500, K-seagateFC530, L-hp900, O-micron3400}) and/or \inum{ii}~minimum bandwidth (by 41.4\%/54.5\% on average across \ssdlist{B-wd850, C-solidigmP41, G-hynixP41, K-seagateFC530, L-hp900}), incurring more severe performance fluctuations than at the early stage.
We also observe that the performance-drop patterns remain largely the same across the various SSD lifetime stages.
Given their consistent peak bandwidth across lifetime stages, this result clearly shows that the performance impact of read disturbance increases with PEC.
Interestingly, \ssdlist{D-samsungPM9A1} and \ssdlist{O-micron3400} provide similar performance regardless of their ages.
We assume that these SSDs may \inum{i}~sustain the reliability of stored data by programming an aged block more precisely (e.g.,~programming data with wider \vth margins can significantly enhance the data's reliability, but at the cost of increased program latency~\cite{jeong-fast-2014, kim-glsvlsi-2018, feng-iccd-2017, wang-tcom-2016, dong-tcas-2010}),
\inum{ii}~perform read-disturbance management conservatively even at early lifetime stages, or \inum{iii}~manufacturers conservatively set the PEC limit (i.e., SSD lifetime) before errors can significantly affect SSD performance.

\take{The performance impact of read disturbance increases with PEC in general, but proactive and conservative reliability management can mitigate such an effect.}

\subsection{Impact of SSD Idle Time}\label{ssec:idle_time_test}
\begin{figure}[h]
     \centering
    \includegraphics[width=\linewidth]{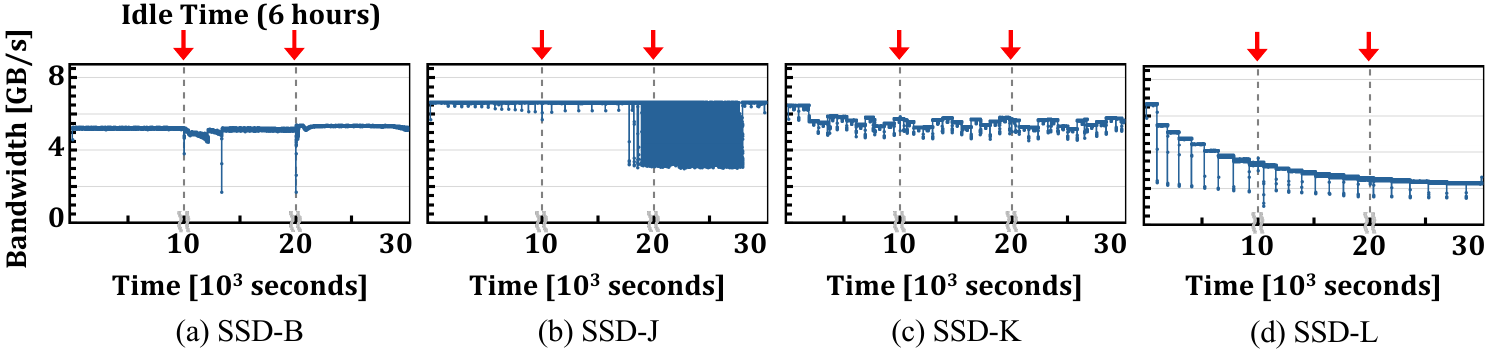}
     \par\addvspace{-5pt}
     \caption{Sequential-read test results with idle times.}
     \label{fig:idle_read}
     \par\addvspace{-5pt}
\end{figure}

We analyze how the performance impact of read disturbance changes in the presence of SSD idle times that can potentially be leveraged for SSD-internal management tasks.
To this end, we repeat the same sequential-read test for all 15 SSDs while providing SSD idle times by issuing no I/O requests for a certain amount of time.
\fig{\ref{fig:idle_read}} shows the read bandwidth of (a) \ssdlist{B-wd850}, (b) \ssdlist{J-crucialT500}, (c) \ssdlist{K-seagateFC530}, and (d) \ssdlist{L-hp900}, when 6-hour idle time (vertical dotted lines) is given for every 10,000~seconds during the sequential-read test for 30,000~seconds with a 1-GB read range.

\observation{obs:idle_time}{Most tested SSDs show largely consistent performance-drop patterns even in the presence of long (six hours) SSD idle times.}
SSD idle times hardly change the performance-drop patterns of \ssdlist{B-wd850, J-crucialT500, K-seagateFC530, L-hp900}; all four SSDs fail to recover or maintain the initial read bandwidth after long idle times.
We observe similar trends for all tested SSDs, which clearly shows that they perform \emph{no} background SSD-internal management tasks for read disturbance.
Doing so leaves the SSDs suffering from performance degradation due to read disturbance, but can avoid other issues potentially caused by background tasks, such as \inum{i}~delaying future user I/Os, \inum{ii}~consuming additional energy, and \inum{iii}~increasing P/E cycles.

\take{Many modern SSDs do \textbf{not} proactively perform read-disturbance management using idle times, contrary to the common assumption that idle times would be utilized for various internal tasks~\cite{ha-tcad-2016, cai-dsn-2015, kang-cm-2017, lee-ispass-2011, chang-vlsi-2016, li-todaes-2021,liao-tecs-2022}.}

\section{Case Study: SSD-Performance Attacks}\label{sec:performance_attack}
To emphasize the system-level performance impact of read disturbance, we showcase a new possible SSD-performance attack.
Our key idea is to deliberately amplify read-disturbance management overheads, enabling an adversary to timely degrade the I/O performance of co-running applications that share the same SSD.
We hypothesize that such a performance attack would be highly feasible, as commodity SSDs \emph{equally} schedule the I/O requests from different processes, regardless of the SSD-internal management overheads for each process~\cite{tavakkol-isca-2018, nvmespec, joshi-hotstorage-2017, hedayati-atc-2019}; 
if an adversary maliciously exacerbates performance overheads, it would also affect the I/O performance of benign applications sharing the same SSD.
There could be other ways simpler and/or more effective to degrade the I/O performance of co-running applications, e.g., issuing massive I/Os with more processes~\cite{yang-sosp-2015, shue-osdi-2012, woo-fast-2021, kwon-fast-2020, huang-fast-2017} or incurring garbage collection frequently via consistent writes~\cite{huang-fast-2017, tavakkol-isca-2018, yan-fast-2017, kim-fast-2015}, but this work is the first to experimentally demonstrate that read disturbance can also be exploited for a storage-related \emph{denial-of-service} (DoS) attack on commodity SSDs.

\subsection{Attack Model} \label{ssec:attack_model}
We assume an adversary only with user-level access to the same SSD as target applications, but no capability to modify SSD firmware or escalate host-system privileges.
We also assume that the host system limits the adversary's SSD access in three aspects.
First, to ensure data privacy, the host dedicates an exclusive LBA range of the shared SSD to each user, e.g., mounting multiple file systems on the SSD, which can avoid the adversary's direct access to any data of target applications.
Second, for performance isolation, the host limits the number of CPU cores available for each user, preventing the adversary from issuing excessive I/Os than other users.
Third, to avoid premature SSD wear-out, the host restricts each user's write traffic, hindering write-based SSD-performance attacks.
For example, when an SSD supports 14,016~TBW (terabytes written) in total with a 5-year warranty~\cite{micron9400-8TB}, the maximum write rate to the SSD is limited to 91~MB/s.

We assume that the adversary can meet the following two key requirements for successfully performing our proposed attack.
First, the attacker can accurately measure the performance of I/O requests in their LBA space, which is necessary to prepare the attack.
Second, the attacker can detect the execution of the target application to timely trigger the attack.

\subsection{Attack Patterns} \label{ssec:attack_pattern}
\begin{figure}[h]
     \centering
     \includegraphics[width=0.78\linewidth]{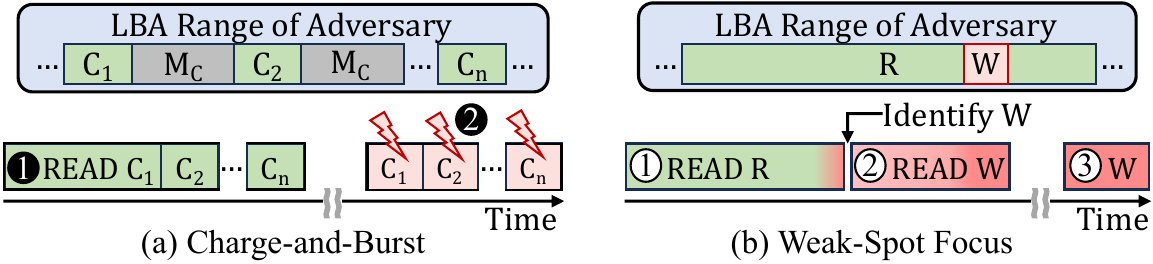}
     \caption{Overview of two attack patterns.}
     \label{fig:worst_cases}
\end{figure}

Based on our observations in \sect{\ref{sec:characterization_results}}, we devise two attack patterns, both of which consist of two phases: \inum{i}~\emph{preparation} and \inum{ii}~\emph{degradation}.
In general, both attacks accumulate read-disturbance to NAND flash blocks during the preparation phase to exaggerate the performance overheads of read-disturbance management techniques inside the SSD during the degradation phase.
We aim to maximize performance degradation during the degradation phase while minimizing the time required for the preparation phase.
\fig{\ref{fig:worst_cases}} presents the high-level overview of the two attack patterns, called (a)~\emph{\AttackA} (CAB) and (b)~\emph{\AttackB} (WSF), respectively.

\head{\AttackA Pattern}
The key idea of \attackA is to \emph{concentrate} the periodic performance drops observed in our sequential-read test into a short time window.
To this end, we \bcirc{1} first accumulate (i.e.,~\emph{charge up}) read disturbance to multiple chunks (e.g.,~C$_k$ in \fig{\ref{fig:worst_cases}(a)}) during the preparation phase, by repeating sequential reads on each chunk (one after the other) until \emph{right before} a performance drop occurs.
During the degradation phase, we \bcirc{2} again repeat per-chunk sequential reads until a significant performance drop occurs and move to the next chunk immediately once the drop is over, which \emph{bursts} performance drops with short intervals.

For \attackA-based attacks, it is crucial to timely terminate the preparation phase for each chunk; 
early termination defers a performance drop during the degradation phase, whereas late termination can cause a premature drop during the preparation phase.
To address this, we employ two termination policies for the preparation phase;
\inum{i}~if an SSD exhibits long drop durations (e.g., \ssdlist{A-samsung980} and \ssdlist{B-wd850}), we stop the preparation phase of each chunk upon a \emph{slight} performance drop;
\inum{ii}~otherwise, we limit the preparation-phase time to the minimum drop period observed in our profiling.
To further reduce premature drops, we also reserve a margin after each chunk (e.g.,~M$_\text{C}$ in \fig{\ref{fig:worst_cases}(a))}, which is large enough to prevent different chunks from being stored in the same NAND flash block.
Doing so ensures that reading a WL disturbs only one chunk's data, thereby avoiding unintended performance drops at C$_k$ while reading C$_{k+1}$ during the preparation phase.

\head{\AttackB Pattern}
The key idea of \AttackB (\attackB) is to \emph{sustain} significant performance drop by reading only a \emph{weak spot}, i.e.,~a data region that causes higher reliability-management overhead than other regions (as in \cref{obs:weak_WLs}).
To this end, during the preparation phase, we \wcirc{1} accumulate read disturbance by sequentially reading a large range of data (e.g., R in \fig{\ref{fig:worst_cases}(b)}) until identifying a weak spot that incurs performance drops (e.g., W~in \fig{\ref{fig:worst_cases}(b))}.
We then \wcirc{2} concentrate sequential reads on the weak spot to accelerate read-disturbance accumulation.
Once the read bandwidth degrades sufficiently, we \wcirc{3} repeat sequential reads only to the weak spot during the degradation phase.

\subsection{Evaluation} \label{ssec:attack_result}
\begin{figure}[h]
     \centering
     \includegraphics[width=0.72\linewidth]{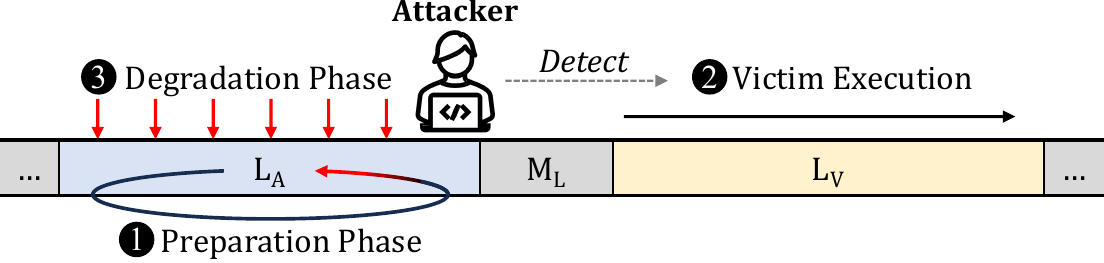}
     \caption{Overview of proposed SSD-performance attack.}
     \label{fig:data_layout}
\end{figure}

\head{Methodology}
\fig{\ref{fig:data_layout}} shows the high-level overview of our methodology to evaluate the effectiveness of the proposed attacks when two processes, attacker and victim, share the same SSD.
Each of the two processes has access only to a contiguous and non-overlapping LBA range of the SSD (e.g.,~L$_\text{A}$ and L$_\text{V}$ in \fig{\ref{fig:data_layout}}).
To avoid unintended read disturbance to the victim's data due to reading of the attacker's data, we set a margin (e.g.,~M$_\text{L}$ in \fig{\ref{fig:data_layout}}) between the two LBA ranges.
The attacker first \bcirc{1} enters the preparation phase, accumulating read disturbance by repeating sequential reads throughout its LBA range.
Once the attacker detects \bcirc{2} the victim's execution, it \bcirc{3} triggers the degradation phase.

We adaptively use the \attackA and \attackB patterns depending on the target SSD's performance-drop characteristics as in \cref{obs:drop_pattern}.
We use the \attackA pattern for SSDs that exhibit steep and periodic performance drops (\ssdlist{A-samsung980, B-wd850, D-samsungPM9A1, G-hynixP41, J-crucialT500, M-solidigmP44}) and the \attackB pattern for SSDs that show consistent performance degradation over time in certain regions (\ssdlist{G-hynixP41, K-seagateFC530, L-hp900, M-solidigmP44}).
We identify that \ssdlist{G-hynixP41} and \ssdlist{M-solidigmP44} meet both conditions, which allows us to use a \emph{combined} pattern; we first profile weak spots throughout a large LBA region and use them as candidate data chunks for a \attackA-based attack.
\tbl{\ref{tab:terminology}} summarizes the sizes of data chunks and ranges used in our evaluation of SSD-performance attacks.
Note that the parameters used in our evaluation may \emph{not} be optimal, as our goal in this case study is to emphasize the importance of better read-disturbance management in modern storage systems, but \emph{not} to develop or optimize the attacks themselves for practical deployment.

\begingroup
\def\arraystretch{1}
\setlength{\tabcolsep}{15pt}
\begin{table}[h]
    \renewcommand\theadfont{\bfseries}
    \caption{Summary of the parameters for the proposed SSD-performance attacks.}
    \label{tab:terminology}
    \centering
    \resizebox{0.7\columnwidth}{!}{
    \small
    \begin{tabular}{ccc}
    \specialrule{\heavyrulewidth}{0pt}{1pt} %
    \thead{Terminology} & \thead{Definition} & \thead{Size}\\
    \specialrule{\heavyrulewidth}{-1.2pt}{2pt} %
        L$_\text{A}$/L$_\text{V}$ & LBA ranges of attacker/victim & 150~GiB\\
        M$_\text{L}$ & Margin between L$_\text{A}$/L$_\text{V}$ & 50~GiB\\
        C$_1$--C$_5$ & Five data chunks used for \attackA & 32~MiB\\
        M$_\text{C}$ & Margin between C$_k$ and C$_{k+1}$ & 5~GiB\\
        R & Large data range used for \attackB & 50~GiB\\
        W & Weak spot used for \attackB & 32~MiB\\
    \bottomrule
    \end{tabular}
    }
\end{table}
\endgroup

\begin{figure}[h]
     \centering
     \includegraphics[width=\linewidth]{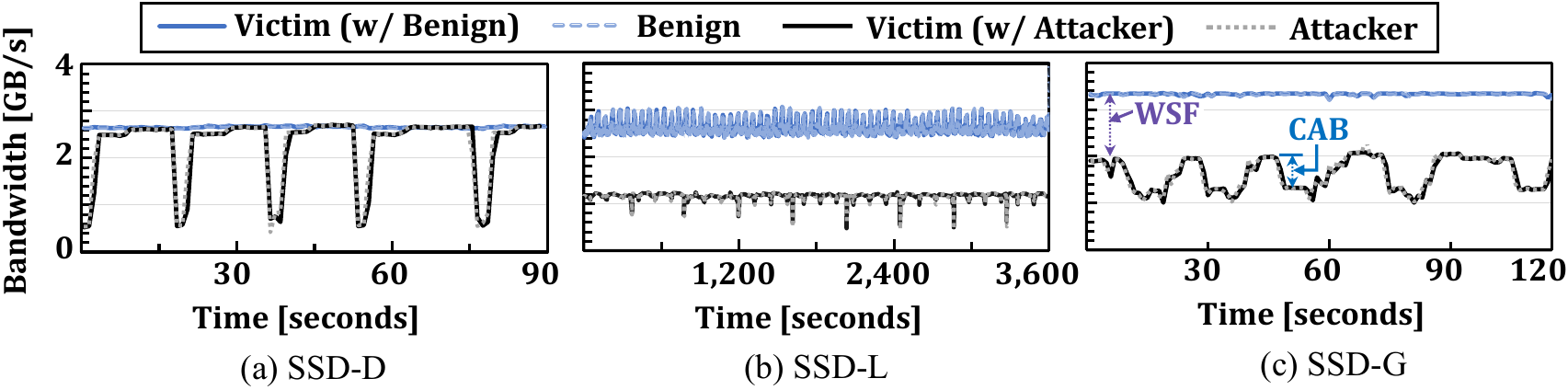}
     \caption{Read bandwidth depending on co-running processes.}
     \label{fig:perf_attack}
\end{figure}

\head{Evaluation Results}
To evaluate the effectiveness of our SSD-performance attacks, we compare the victim's read bandwidth under two execution scenarios: co-running with \inum{i}~the attacker and \inum{ii}~another benign process.
Both the victim and benign processes sequentially read their exclusive LBA range (150~GiB) at the pristine state, i.e., no read disturbance accumulated.
To assess the proposed attack's timeliness, we separate its preparation and degradation phases with an SSD idle time ($>$ 1~hour).
\fig{\ref{fig:perf_attack}} shows the most representative results for (a)~\attackA-based attack (\mbox{\ssdlist{D-samsungPM9A1}}), (b)~\attackB-based attack (\ssdlist{L-hp900}), and (c) combined attack (\ssdlist{G-hynixP41}) from our evaluation of all 15 tested SSDs at their late lifetime stages.
We repeat each experiment three times and confirm that the proposed performance attacks successfully and considerably degrade the victim's read bandwidth in most tested SSDs: \ssdlist{A-samsung980, B-wd850, D-samsungPM9A1, G-hynixP41, J-crucialT500, M-solidigmP44} (\attackA), \ssdlist{G-hynixP41, K-seagateFC530, L-hp900, M-solidigmP44, O-micron3400} (\attackB), and \ssdlist{G-hynixP41, M-solidigmP44}(combined).

\observation{obs:attack_individual}{Both our \attackA- and \attackB-based attacks can significantly degrade the I/O performance of the victim process, requiring neither reading the victim's data nor issuing writes.}
As shown in \fig{\ref{fig:perf_attack}(a)}, the \attackA-based attack on \ssdlist{D-samsungPM9A1} causes significant performance drops five times (i.e., at all chunks C$_1$ to C$_5$) within 90~seconds, each degrading the victim's read bandwidth by 74.8\% for 3~seconds on average.
Given that the average drop period observed during our entire sequential-read tests on \ssdlist{D-samsungPM9A1} is $>$2,000~seconds, the result clearly demonstrates the \attackA-based attack's capability to burst performance drops.
It is possible to induce more performance drops by using more chunks, but this comes at the cost of longer preparation phase.

The \attackB-based attack also significantly affects the victim's I/O performance, but in a different way from the \attackA-based attack.
As shown in \fig{\ref{fig:perf_attack}(b)}, the attack on \ssdlist{L-hp900} \emph{consistently} degrades the read bandwidth of the victim process by 54.2\% on average during the entire degradation phase that we limit to 3,600 seconds.
We expect that the \attackB-based attack can affect the victim's I/O performance even for several hours regardless of idle time, given that all \typeB SSDs cannot recover the read bandwidth of weak spots during the sequential-read test for 30,000 seconds.

\observation{obs:attack_combined}{The \attackA and \attackB patterns can be synergistically combined to devise a more severe SSD-performance attack that further degrades the victim's I/O performance.}
As shown in \fig{\ref{fig:perf_attack}(c)}, using the weak spots prepared by the \attackB pattern enables the attacker process to limit the victim's read bandwidth to around 2~GB/s (40.9\% lower than the no-attack baseline) during the degradation phase.
In the meantime, the attacker also successfully causes further performance drops by 70.4\% at all five weak spots with the \attackA pattern.
Note that the attacker can skip the \attackB pattern's preparation phase for future combined attacks as long as the used weak spots remain degraded ($\geq$ 30,000 seconds in our sequential-read test).

\observation{obs:attack_scheduling}{Two concurrent processes show almost \emph{identical} performance in all three SSDs.}
As shown in \fig{\ref{fig:perf_attack}}, the victim's read bandwidth closely tracks that of the concurrently running process (attacker or benign) with minimal deviation. 
We observe the same trend in all 15 tested SSDs, which validates our hypothesis; 
the common I/O-scheduling policy in modern SSDs makes our proposed performance attacks highly feasible by \emph{unfairly} forcing the victim process to suffer half of the attacker-induced slowdown.

\take{Under the common I/O-scheduling policy in modern SSDs, an application's I/O performance can degrade when it shares the same SSD with other read-intensive applications, much more significantly than expected.}

\section{Macro-Benchmark Evaluation}\label{sec:macro_benchmark}
Although our primary goal in this work is to analyze the system-level performance impact of read disturbance \emph{when it is substantial} (e.g., under heavily read-dominant workloads), we also evaluate read–write mixed macro-benchmarks to understand the impact in more general cases.

\subsection{Methodology}\label{ssec:filebench_methodology}
\head{Workloads}
We use the \webserver and \videoserver workloads from the FileBench benchmark tool~\cite{filebench} with the direct I/O option, both maintaining separate file sets for read and write operations.
The \webserver workload consists of multiple threads, each of which randomly reads several files from the data-file set and then appends data to a single shared log file.
The \videoserver workload comprises two types of threads: \inum{i}~multiple reader threads, which randomly read files from the active file set, and \inum{ii}~a single writer thread, which periodically deletes a file from the passive file set and writes a new file to it.

\tbl{\ref{tab:filebench_config}} summarizes the key characteristics of the workloads used in our evaluation.
We adjust each workload to vary the read-write ratio by configuring \inum{i}~the number of files read per log update to five (write-heavy) and ten (read-heavy) for \webserver and \inum{ii}~the file-replacement interval to one second (write-heavy) and ten seconds (read-heavy) for \videoserver.
The default \webserver workload monotonically increases the log file's size, quickly exhausting the entire SSD capacity and thus preventing long-term evaluation.
To address this, we introduce a log-backup process that truncates the entire log file after copying it to another storage device periodically (every 15 and 20 minutes for write-heavy and read-heavy workloads, respectively).

\begin{table}[h]
  \caption{FileBench workload characteristics.}
  \label{tab:filebench_config}
  \centering
  \setlength{\tabcolsep}{9pt}
  \resizebox{\columnwidth}{!}{%
    \begin{tabular}{cccccccc}
      \toprule
      \multirow{2}{*}{\textbf{Workload}} &
      \multirow{2}{*}{\makecell{\textbf{File}\\\textbf{set type}}} &
      \multirow{2}{*}{\makecell{\textbf{Average}\\\textbf{file size}}} &
      \multirow{2}{*}{\makecell{\textbf{Number}\\\textbf{of files}}} &
      \multicolumn{2}{c}{\textbf{I/O sizes}} &
      \multirow{2}{*}{\makecell{\textbf{Number of}\\\textbf{threads}}} &
      \multirow{2}{*}{\makecell{\textbf{Read/Write}\\\textbf{ratio}}} \\
      \cmidrule(lr){5-6}
      & & & & \textbf{Read} & \textbf{Write} & & \\
      \midrule
      \multirow{2}{*}{\textbf{\textsf{Web Server}}}
        & Data & \multirow{2}{*}{16~KB} & 65,536 & 16~KB & --    & \multirow{2}{*}{100}
        & \multirow{2}{*}{\makecell{WH$^{1}$: 5:1\\RH$^{2}$: 10:1}} \\
        & Log  &                       & 1      & --    & 16~KB &                      & \\
      \addlinespace
      \multirow{2}{*}{\textbf{\textsf{Video Server}}}
        & Active  & \multirow{2}{*}{1~GB} & 10  & 1~MB & --   & 33
        & \multirow{2}{*}{\makecell{WH: 4.6--40:1\\RH: 33--44:1}} \\
        & Passive &                       & 432 & --   & 1~MB & 1              & \\
      \bottomrule
    \end{tabular}%
  }
  \vspace{2pt}
  {\footnotesize\raggedright{$^{1}$WH: Write-heavy,\quad $^{2}$RH: Read-heavy.}\par}
\end{table}

\head{Preconditioning}
We precondition both the file system and SSD before each FileBench experiment in three steps to reduce potential distortions caused by previous device and file system states. 
First, we delete all files under the target mount point to reset the file-system namespace. 
Second, we issue a TRIM command to the mount point to inform the SSD of invalidated LBAs via the \texttt{fstrim} command. 
Third, we disable address space layout randomization (ASLR), as some SSDs do not operate reliably under randomized memory layouts during FileBench test.
All experiments run on the ext4 file system, and we monitor SSD performance using the Linux \texttt{iostat} tool.

\subsection{Evaluation Results}\label{ssec:filebench_results}
\begin{figure}[h]
     \centering
     \includegraphics[width=\linewidth]{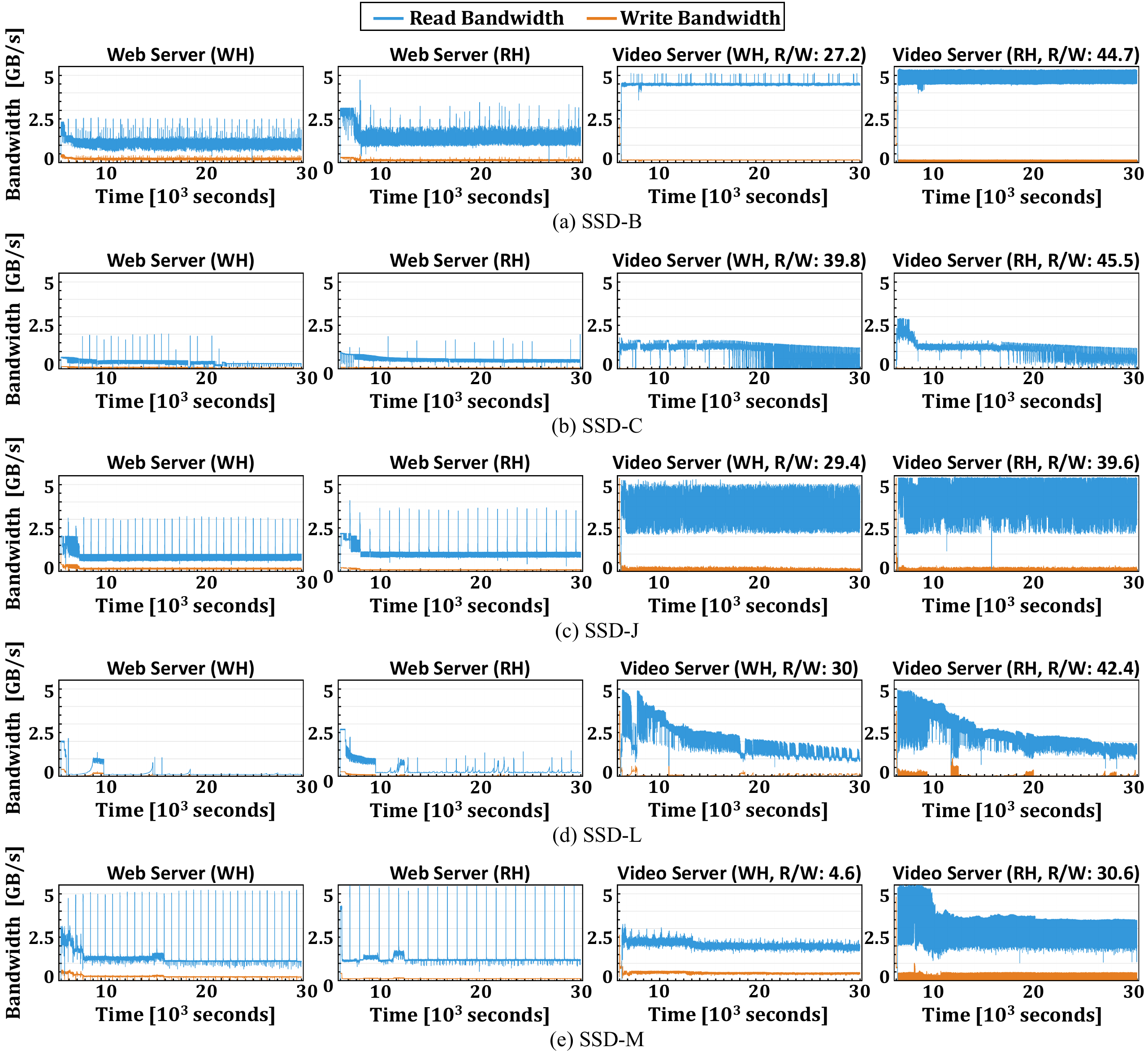}
     \caption{FileBench test results.}
     \label{fig:filebench_result}
     \par\addvspace{-10pt}
\end{figure}

\fig{\ref{fig:filebench_result}} shows the read and write bandwidth of (a)~\ssdlist{B-wd850}, (b)~\ssdlist{C-solidigmP41}, (c)~\ssdlist{J-crucialT500}, (d)~\ssdlist{L-hp900}, and (e)~\ssdlist{M-solidigmP44} during FileBench test for 30,000~seconds.
For the \videoserver workloads, we denote the actual read-write ratio (\textsf{R/W}), which varies depending on the SSD's request scheduling policy;
unlike in \webserver where each thread performs both read and write operations, threads in \videoserver perform either read or write operations only, so the amounts of reads and writes actually performed highly depend on how the SSD schedules I/Os from different threads. 
Note that most tested SSDs exhibit periodic read-bandwidth spike in \webserver, which occurs during the log-backup process, i.e.,~sequential reading of the entire log file \emph{without} read disturbance or write-induced interference.

\observation{obs:fb_less_impact}{In many cases, the performance impact of read disturbance becomes less severe under read-write mixed workloads compared to read-only workloads.}
\ssdlist{B-wd850} exhibits only a single severe performance drop that we suspect is caused by read disturbance, i.e., near-zero read-bandwidth lasting 2~seconds at 21,236~seconds under the read-heavy \webserver workload, which is too infrequent to be attributed to garbage collection. 
Under all FileBench workloads, \ssdlist{J-crucialT500} does not exhibit the significant performance fluctuations observed under the read-only workloads.
Similarly, under the \webserver workloads issuing a number of small random reads, \ssdlist{L-hp900} and \ssdlist{M-solidigmP44} do not experience the performance fluctuations observed under the random-read tests (cf. \figs{\ref{fig:rand_1GB}(l) and \ref{fig:rand_1GB}(m)}).

The results are largely expected and can be readily explained by the fact that writes are well known to affect SSD performance more severely than reads, due to the long program latency of NAND flash memory and various write-related FTL tasks, such as garbage collections and wear leveling.
Write-induced degradation of overall I/O bandwidth can decrease the performance impact of read disturbance in two ways.
First, the substantial performance impact of writes dominates overall SSD performance, thereby making the relative impact of read disturbance less pronounced.
Second, it slows down the accumulation of read disturbance, which, in turn, reduces the frequency of read-disturbance-induced reliability management tasks.

\observation{obs:fb_remaining_impact}{In some SSDs, read disturbance can have a non-trivial impact on I/O performance under read-write mixed workloads.}
We observe that \ssdlist{C-solidigmP41} exhibits consistent peak-performance degradation under all FileBench workloads, unlike the behavior we observe under read-only workloads, where it can periodically recover its peak bandwidth (cf. \figs{\ref{fig:seq_1GB}(c) and \ref{fig:rand_1GB}(c)}).
In particular, significant performance fluctuations occur after 20,000~seconds under the \videoserver workloads, which are also not observed under any read-only workloads.
As explained in \sect{\ref{ssec:read_range_test}} (\cref{obs:seq_readrange_exception}), we hypothesize that \ssdlist{C-solidigmP41} proactively migrates read-hot data to other physical blocks; unfortunately, such migration becomes difficult in the presence of user writes, causing accumulated read disturbance to consistently affect SSD performance.
We also observe continuous performance drops over time in \ssdlist{L-hp900} and \ssdlist{M-solidigmP44} under the \videoserver workloads, similarly under sequential read workloads (cf. \figs{\ref{fig:seq_1GB}(l) and \ref{fig:seq_1GB}(m)}).
Under the read-heavy (write-heavy) \videoserver workload, \ssdlist{L-hp900} and \ssdlist{M-solidigmP44} exhibit considerable read-bandwidth reductions of 55.9\% (57.8\%) and 22.5\% (19.8\%), respectively, when comparing the last five minutes to the first five minutes.
Note that the performance drops in \ssdlist{L-hp900} and \ssdlist{M-solidigmP44} under \videoserver differ from those observed in \ssdlist{B-wd850, J-crucialT500, L-hp900, M-solidigmP44} under \webserver.
We attribute the performance drops under \webserver to SLC-buffer flushes, as the workloads initially write only a limited amount of data ($\approx$1~GB).
In contrast, the \videoserver workloads rapidly exhaust the SLC buffer at the beginning by sequentially writing the large passive file set, making it plausible that read disturbance is the root cause of the continuous performance drops.

\take{The performance impact of read disturbance is generally reduced in the presence of write requests, but read disturbance can still introduce non-trivial performance degradation in some operating scenarios (SSDs and workloads).}

\section{Implications} \label{sec:implication}
Our observations made in \sect{\ref{sec:characterization_results}} and \sect{\ref{sec:performance_attack}} can guide host- and SSD-side future improvements for modern storage systems to enhance I/O performance for read-intensive applications.
We leave the development of new techniques for such improvements to future work, which can benefit from not only our guidance but also observations that we make on the behavior of read-disturbance management techniques applied in commodity SSDs (e.g., for developing a more accurate SSD simulator that can model more realistic SSD performance).

\subsection{Host-Side Improvements}
Our characterization study can help system designers and application developers improve the I/O performance of their systems and applications, respectively, in two ways.

\head{Better Understanding of System I/O Performance}
Observations \Cref{obs:seq_specification} to \Cref{obs:rand_specification} demonstrate that the \emph{actual} SSD performance under read disturbance varies across SSDs and differs from vendor specifications. 
A deeper understanding of such actual performance characteristics can allow more efficient system and application design in two ways.
First, system designers can better provision storage resources to meet their performance requirements at minimal cost, selecting the appropriate SSD model and quantity to ensure that storage systems offer the desired performance under the expected read patterns of critical applications.
Second, application developers can better coordinate tightly coupled compute and data-access tasks.
For example, a large language model (LLM) inference framework that relies on timely I/O prefetching~\cite{sheng-icml-2023, alizadeh-arxiv-2023, aminabadi-arxiv-2022} can achieve higher effectiveness through precise pipelining based on more accurate performance estimation.

\head{Host-SSD Interaction for Better Read-Disturbance Management}
Observations~\Cref{obs:weak_WLs}, \Cref{obs:idle_time}, and \Cref{obs:attack_individual} show that read disturbance-induced I/O slowdown may remain indefinitely even with sufficient SSD idle times.
To recover I/O performance for critical data, an application itself needs to rewrite the data (after reading it) to the SSD.
Even though the application can accurately determine the timing and target data for read reclaim, such application-managed read-disturbance management can cause non-trivial inefficiencies due to costly host-SSD data movements.
Such inefficiencies can be addressed by introducing a \emph{new storage interface} that enables the host to proactively trigger read reclaim on target data, similarly to NVMe I/O determinism (IOD)~\cite{nvmespec-iod}.
Currently, IOD allows the host to temporarily forbid SSD-internal management tasks, but it offers no mechanism for the host to proactively trigger those tasks or specify their target data.
We expect that the development of such an interface can accurately inform SSDs of when read reclaim is needed and be further extended to allow the host to specify data to be stored in a more reliable region which is less susceptible to read disturbance (e.g., SLC/MLC blocks or robust WLs that provide better reliability).

\subsection{SSD-Side Improvements} 
Our characterization study can be used in four ways for developing efficient FTL algorithms to better manage read disturbance inside SSDs.

\head{Proactive Read-Disturbance Management}
Observations~\Cref{obs:drop_pattern}, \Cref{obs:seq_readrange_exception}, and \Cref{obs:seq_pec} highlight the importance of proactive read-disturbance management.
As observed, excessive delay of read reclaim can rather cause significant long-term performance degradation or severe performance fluctuation when multiple blocks require read reclaim simultaneously, contrary to common practice of deferring read reclaim until truly necessary to minimize its high impact on application performance~\cite{liu-asplos-21}.
In particular, the effectiveness of proactive management can be further improved by taking into account the lifetime of blocks; increasing proactivity for aged blocks can effectively mitigate read-retry and ECC overheads that worsen at high PEC.

\head{Mitigating Performance Fluctuation due to Read-Disturbance Management}
Observations~\Cref{obs:drop_pattern},~\Cref{obs:idle_time}, and~\Cref{obs:attack_individual} suggest three directions to avoid severe performance drops due to proactive read-disturbance management.
First, we can minimize the performance impact of proactive management by efficiently utilizing over-provisioned parallelism while prioritizing user I/Os.
Second, leveraging SSD idle time can possibly eliminate user-experienced performance drops.
Third, read-disturbance management should be performed in a \emph{progressive} manner, i.e., reclaiming a few blocks preemptively rather than reclaiming a number of blocks at once.

\head{Adaptive Read-Disturbance Management}
Observations \Cref{obs:rand_less_impact} and \Cref{obs:seq_readrange_exception} suggest that read-disturbance management should take workload I/O patterns into account in two aspects. 
First, read reclaim should be prioritized for data accessed sequentially over data referenced randomly since read disturbance has less impact under random reads compared to sequential reads.
Second, promoting \textit{read-intensive data} to SLC or MLC regions, which are less susceptible to read disturbance, can effectively mitigate the performance overhead of read-disturbance management.
Such promotion must be performed carefully, as it involves substantial data movements across different regions, which can negatively affect SSD performance as shown in \fig{\ref{fig:seq_readrange_comp}}(b).

\head{Read Disturbance-Aware Internal I/O Scheduling}
Observations~\Cref{obs:attack_individual} to \Cref{obs:attack_scheduling} suggest that the SSD controller should isolate internal I/O traffic related to read-disturbance management to minimize their interference with other processes. 
Most SSDs prioritize fairness by equally scheduling I/O requests from concurrently running processes, without distinguishing read disturbance-related reads and writes.
As a result, I/O requests from other users are often delayed, leading to significant performance degradation.
For better performance isolation, the FTL should be able to track the processes that trigger read-disturbance management and impose such overheads only on the corresponding processes.

\section{Related Work}\label{sec:related_work}
To our knowledge, this work is the first to rigorously characterize the system-level performance impact of read disturbance in modern SSDs under various operating conditions, which introduces new observations and takeaways related to read-disturbance management techniques, e.g., their impact depending on workloads, idle times, and co-running processes.
We provide a brief review of prior studies on read-disturbance characterization and mitigation.

\head{Read-Disturbance Characterization}
In addition to the prior work~\cite{liu-asplos-21} that we already discussed in \sect{\ref{sec:motivation}}, many other previous studies have also investigated the reliability impact of read disturbance in NAND flash memory under various multi-leveling techniques, including MLC~\cite{cai-dsn-2015}, TLC~\cite{ren-nvmsa-2023, chun-tos-2024, zambelli-irps-17}, and QLC~\cite{yang-icsict-2024}.
Although these studies conduct extensive analyses of read disturbance, their scope is limited to the device level rather than the system level.
In contrast, our work characterizes the system-level performance impact of read disturbance, providing deeper insights for designing both host systems and SSD architectures to develop more effective read-disturbance management.

\head{Read-Disturbance Mitigation}
Prior studies~\cite{cai-dsn-2015,ha-tcad-2016} have proposed device-level techniques to mitigate the reliability impact of read disturbance.
Cai et al.~\cite{cai-dsn-2015} propose a new page-read mechanism of NAND flash memory that uses a lower \vpass level to reduce the \vth shift caused by read disturbance.
Even though the lower \vpass level slightly increases RBER for the highest \vth state (e.g., P7 in TLC NAND flash memory), its benefit (i.e., reducing read disturbance) outweighs the cost, thereby leading to an overall RBER improvement.
Ha et al.~\cite{ha-tcad-2016} propose to narrow the \vth window of a block via more precise programming, allowing the use of a lower \vpass level without RBER increases for the highest \vth state.
They minimize the performance penalty of precise programming by selectively storing only read-intensive data to more robust blocks with a narrower \vth window.

\head{Read-Reclaim Optimization}
Several prior studies~\cite{ha-tcad-2016, zhang-tcad-2022, han-vlsi-2023,chun-cal-2025,lee-micro-2025} have proposed various firmware-level techniques to mitigate the overheads of read reclaim, the reliability-management task directly related to read disturbance.
Ha et al.~\cite{ha-tcad-2016} propose a new data-placement technique that places read-intensive data to pages that require fewer sensing operations (e.g., LSB/MSB pages in \fig{\ref{fig:flash_reliability}}), which effectively mitigates read disturbance per page read, thereby reducing read-reclaim invocations.
Zhang et al.~\cite{zhang-tcad-2022} also introduce another data-placement scheme that mixes read-hot and read-cold data within the same block to avoid concentrating reads on the specific blocks and thus reduces read reclaims triggered by the same number of page reads.
Han et al.~\cite{han-vlsi-2023} employ different read-count thresholds for read reclaim depending on the page types, which enables a block to service more page reads before migrating the block's disturbed data.
Two recent studies aim to minimize data migration via more fine-grained read reclaim at low cost;
Chun et al.~\cite{chun-cal-2025} leverage the Space-Saving algorithm to efficiently estimate the read disturbance accumulated to each WL, and Lee et al.~\cite{lee-micro-2025} optimize the patrol-read mechanism to accurately monitor the RBER of WLs with minimal performance overheads.

\head{Other Optimizations for Reliability-Management Techniques}
A large body of prior work has proposed various techniques to optimize read retry and ECC, which can also mitigate the performance impact of read disturbance.
First, many studies have extensively optimized the read-retry mechanism.
Shim et al.~\cite{shim-micro-2019} propose a new read mechanism that reuses \vref levels recently used for neighboring WLs to reduce the number of read retries, which we assume are used in \ssdlist{G-hynixP41} and \ssdlist{M-solidigmP44}.
Park et al.~\cite{park-asplos-2021} develop a new read-retry mechanism that reduces the latency of a read-retry operation by leveraging the large ECC margin and the cache-read feature.
Chun et al.~\cite{chun-hpca-2024} introduce an on-die read-retry mechanism that performs read retry inside the chip without ECC decoding, thereby significantly improving effective utilization of internal-channel bandwidth.
Ye et al.~\cite{ye-asplos-2024} try to predict optimal \vref value based on \inum{i}~page type, \inum{ii}~P/E cycles, and \inum{iii}~retention time.
Second, some studies~\cite{li-taco-2024,cui-todaes-2022,liu-ojcas-2022} have proposed various ECC algorithms and hardware decoder designs to enhance error-correction capability and decoding throughput.

Our work is aligned with these efforts but goes a step further in three key aspects.
First, although prior studies have achieved substantial improvements in mitigating the performance impact of read disturbance, our experimental studies using commodity SSDs demonstrate that read disturbance is a fundamental problem that is difficult to completely eliminate.
Second, we identify the potential risk of DoS attacks that exploit read disturbance as a vulnerability and experimentally demonstrate their impact on modern SSDs, which highlights the need to develop new techniques to mitigate performance interference caused by read disturbance in multi-tenant scenarios.
Third, based on our characterization results, we present new optimization directions at the host and interface levels.

\section{Conclusion}\label{sec:conclusion}

In this study, we investigate the system-level performance impact of read disturbance in modern SSDs, which remains largely uninvestigated by prior studies. Through extensive experiments on 15 NVMe SSDs, we demonstrate how significantly read disturbance can degrade I/O performance and even be exploited for potential DoS attacks. Our \minorm{16} observations and \minorm{7} takeaway lessons lead to six key directions for improving host-system and SSD-architecture design, emphasizing the need for more integrated and proactive read-disturbance management.
We hope that the novel experimental results and insights of our study will inspire and aid future work to develop more effective read-disturbance management techniques.

\begin{acks}
We thank our anonymous reviewers of SIGMETRICS 2026 for their valuable feedback and comments.
This work was supported by the National Research Foundation of Korea (RS-2023-00283799, RS-2024-00396850, RS-2024-00415602, and RS-2025-00519994), Institute for Information \& Communications Technology Planning \& Evaluation (RS-2024-00347394 and RS-2024-00437866), and Samsung Electronics Co., Ltd (IO230411-05858-01).
Jisung Park is the corresponding author.
\end{acks}

\bibliographystyle{unsrt}
\bibliography{refs}

\end{document}